\documentclass[amsmath,amssymb,showpacs,showkeywords,twocolumn]{revtex4}
\usepackage{graphicx}
\usepackage{times}
\usepackage{mathrsfs}
\usepackage{amsmath}
\usepackage{graphicx}
\usepackage{epsfig}
\usepackage{dcolumn}
\usepackage{bm}
\usepackage{multirow}
\usepackage[utf8x]{inputenc}
\DeclareUnicodeCharacter{2212}{-}
\begin{document}
\title{Effect of interfacial Dzyaloshinskii - Moriya interaction in spin dynamics of an Antiferromagnet coupled Ferromagnetic double - barrier Magnetic Tunnel Junction}
\author{Reeta Devi\footnote{reetadevi@dibru.ac.in},   Nimisha Dutta\footnote{nimishadutta@dibru.ac.in}, Arindam Boruah\footnote{arindamboruah@dibru.ac.in} and Saumen Acharjee\footnote{saumenacharjee@dibru.ac.in}}
\affiliation{Department of Physics, Dibrugarh University, Dibrugarh 786 004, 
Assam, India}

\begin{abstract}
In this work, we have studied the spin dynamics  of a synthethic Antiferromagnet (SAFM)$|$Heavy Metal (HM)$|$Ferromagnet (FM) double barrier magnetic tunnel junction (MTJ) in presence of Ruderman - Kittel - Kasuya - Yoside interaction (RKKYI), interfacial Dzyaloshinskii - Moriya interaction (iDMI), N\'eel field and Spin-Orbit Coupling (SOC) with different Spin Transfer Torque (STT). We employ Landau-Lifshitz-Gilbert-Slonczewski (LLGS) equation to investigate the AFM dynamics of the proposed system. We found that the system exhibits a transition from regular to damped oscillations with the increase in strength of STT for systems with weaker iDMI than RKKYI while display sustained oscillatons for system having same order of iDMI and RKKYI. On the other hand the iDMI dominating system exhibits self-similar but aperiodic patterns in absence of N\'eel field.   In the presence of N\'eel field, the RKKYI dominating systems exhibit chaotic oscillations for low STT but display sustained oscillation under moderate STT. Our results suggest that the decay time of oscillations can be controlled via SOC. The system can works as an oscillator for low SOC but display nonlinear characteristics with the rise in SOC for systems having weaker iDMI than RKKYI while an opposite characteristic are noticed for iDMI dominating systems. We found periodic oscillations under low external magnetic field in RKKYI dominating systems while moderate field are necessary for sustained oscillation in iDMI dominating systems. Moreover, the system exhibits saddle-node bifurcation and chaos under moderate N\'eel field and SOC with suitable iDMI and RKKYI. In addition, our results indicate that the magnon lifetime can be enhanced by increasing the strength of iDMI for both optical and acoustic modes .
\end{abstract}
\maketitle
\section{Introduction}
Recently, there has been a resurgence of interest in Antiferromagnets (AFM) within the field of spintronics \cite{jungwirth,han,baltz,zelezny,marti,park,wadley,dutta,jungfleisch,cheng,kampfrath,du,li22,kim}. This renewed attention is due to the unique characteristics of AFMs, such as their high resonance frequency in the terahertz range \cite{cheng,kampfrath}, the absence of stray magnetic fields \cite{du}, and their remarkable stability under magnetic fields \cite{baltz, zelezny}. Consequently, devices based on AFMs offer the potential for faster operation compared to traditional ferromagnetic (FM) devices, making them promising candidates for applications in data storage and information processing \cite{jungfleisch}.  Moreover, the recent discovery of electrical switching in AFM based devices via Spin-Orbit Torque (SOT) demonstrates that AFMs can be manipulated electrically in similar ways to their FM counterparts \cite{kim,li22,dutta}. This discovery has sparked significant research interest in AFM spintronics \cite{gomonay,kosub2,olejnik,lebrun,chen}. With further discovery of Gaint Magnetoresistance (GMR) \cite{baibich} and Spin transfer Torque (STT) \cite{slonczewski1,slonczewski2,berger} in magnetic tunnel junctions (MTJ) the AFM based heterostuctures and magnetic random access memories (MRAM) received significant boost as a material for future technology.

A typical MTJ usually consists of a tunnel barrier between two ferromagnetic layers,  act as pinned and free layers. However, such configurations face issues with thermal stability below 40 nanometers \cite{li2}. To overcome this, researchers have turned into double-interface MTJs, which involve placing a heavy metal (HM) layer between two ferromagnetic layers \cite{choi,sato,li2,acharjee91,garzon,lee99,iwata,parkin}. This FM$|$HM$|$FM configuration offers better thermal stability and plays a crucial role in enhancing spin-orbit coupling (SOC) and also in generation of the Ruderman-Kittel-Kasuya-Yosida interaction (RKKYI) \cite{garzon,lee99,iwata}. It is to be noted that the  RKKYI ferromagnetically couple the magnetizations of the two layers, resulting them behave like identical layers \cite{parkin, li2, acharjee91}. Additionally, the lack of inversion symmetry in these systems can  also generate an anti-symmetric interfacial Dzyaloshinskii-Moriya interaction (iDMI) which chirally couple the spins \cite{dzyaloshinsky, moriya,cho,caretta,pacheco,rakibul}. Moreover, the emergence of iDMI can also be triggered via the strong SOC of the HM layer and hence play significant role in the formation of magnetic textures, such as chiral domains \cite{ ding}, magnetic skyrmions \cite{yu99,wolf}, and N\'eel-type domain walls \cite{park91}. 
Recent studies suggest that the RKKYI counteract the adverse effects of iDMI in STT-induced switching \cite{li2,acharjee91,parkin}.  Consequently, it is important to comprehend how SOC, RKKYI, and iDMI collectively influence the STT-induced spin dynamics of AFM based MTJs. 

AFM based MTJs require low STT to switch different resistance states \cite{zink, zhao91} and also has improved thermal stability \cite{rozsa, iwata}. Thus, these devices are more energy efficient and suitable for high temperature operations. Moreover, this features also enable higher data storage density in MRAM \cite{iwata, yang91}. Apart from that AFM based MTJs have potential applications in spin-transfer oscillators for microwave signal generation due to their low STT and stability \cite{volvach}. Numerous studies have been done to investigate the mechanism behind SOT-induced N\'eel vector switching and to gain a better understanding of the AFM dynamics in various hybrid structures considering N\'eel SOT \cite{wu91,zhang92,chiang,xu51,gomonay2,gomonay3,yuan3}. Efforts have been made to comprehend the roles of DMI and field free SOT switching in synthetic antiferromagnets (SAFM) \cite{chen91}. It is to be noted that the future development and application of AFM based MTJs rely on our comprehensive understanding of spin dynamics of the AFM order in AFM based MTJs where AFM is considered as a storage layer. However, till present the impact of RKKYI and iDMI on AFM dynamics has not been explored. Furthermore, the influence of SOC, STT and N\'eel field on spin dynamics have not been considered within the same framework in prior research. So we investigated the effect on iDMI, RKKYI, SOC and N\'eel field on spin dynamics of a double barrier SAFM$|$HM$|$FM based MTJ in presence of STT. 

The organization of this paper are as follows: In Section II, we present a minimal theory to study the time evolution of the AFM and FM order of the proposed system. The results of our work is presented in Section III, where we consider the impact of RKKYI, iDMI, RSOC, N\'eel field and other crucial parameters like external magnetic field and STT on spin dynamics.  Additionally, in this section, we explore the influence of iDMI on the lifetime and stability of the magnons in the system. We conclude with a concise summary of our work in Section IV.

\section{Minimal Theory}
\subsection{Time evolution of the AFM and FM order parameters}
\begin{figure}[hbt]
\centerline
\centerline{
\includegraphics[scale=0.75]{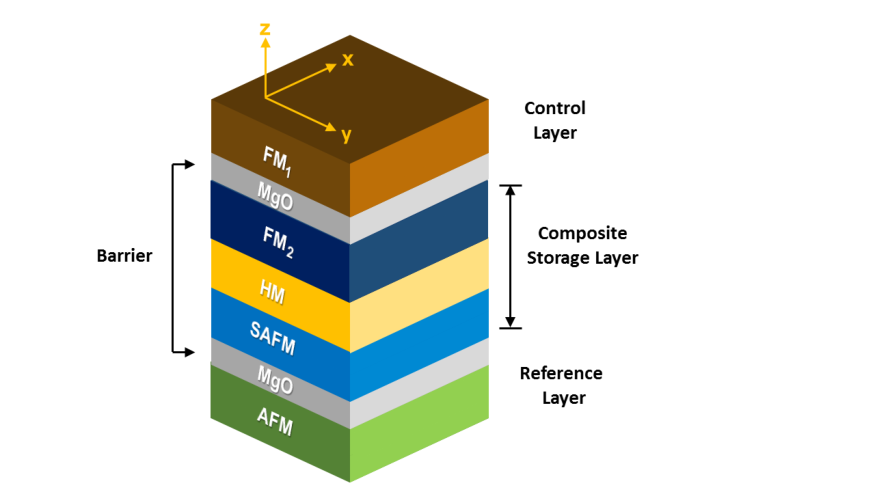}
\vspace{-0.1cm}
}
\caption{Schematic representation of the proposed double barrier Antiferromagnet (AFM) coupled Ferromagnetic (FM) MTJ. A free FM$_2$ coupled to a Synthethic Antiferromegnet (SAFM) through a Heavy Metal (HM) is considered as a composite storage layer. The top FM$_1$ layer is the control layer while the bottom  AFM layer is considered as a reference layer for read-write operation in MTJ.
}
\label{fig1}
\end{figure}

The schematic illustration of a double - barrier AFM coupled FM MTJ is shown in Fig. \ref{fig1}. A typical double - barrier MTJ consist of three magnetic layers viz. reference layer, storage layer and control layer. The reference and control layers work as polarisers whose polarizations can be controlled autonomously of the free layer \cite{acharjee92}. We consider an AFM reference layer and an FM$_1$ control layer with SAFM$|$HM$|$FM$_2$ composite free layer for our analysis. This unconventional SAFM$|$HM$|$FM$_2$ composite layer can have several advantages in controlling and tuning the magnetization vector via  SAFM layer through STT and SOT. Also, the unconventional pairing of SAFM with FM$_2$ via HM can result in asymmetric exchange coupling like interfacial Dzylashinskii - Moriya interaction (iDMI), Ruderman - Kittel - Kasuya - Yoside interaction (RKKYI)  in the proposed MTJ. Moreover, an STT is induced in the storage layer as the polarized current passes through it \cite{li2, acharjee91, parkin, coelho}. An easy axis anisotropy along the z-direction and an external magnetic field along the x-direction is considered in our analysis. 

Classically, an FM can be described by the magnetization $\mathbf{m}$ while two sublattice AFM can be described by order parameters $\mathbf{m}_1$ and $\mathbf{m}_2$. The AFM N\'eel order parameter can be redefined as $\mathbf{n} \equiv \mathbf{m}_1 - \mathbf{m}_2$ and the FM order parameter as $\mathbf{m} \equiv \mathbf{m}_1 + \mathbf{m}_2$. The time evolution of the FM and AFM order parameters can be studied by using coupled Landau - Lifshitz - Gilbert - Slonczewski (LLGS) equations \cite{gomonay2,gomonay3,acharjee91}. 
\begin{align}
\label{eq1}
\mathbf{\dot{m}} &=-\mathbf{m}\times \left(\gamma\mathbf{H}_\text{m}-\alpha _\text{m}\mathbf{\dot{m}}\right)-\mathbf{n}\times \left(\gamma\mathbf{H}_\text{n}-\alpha _\text{n}\mathbf{\dot{n}}\right)+\mathbf{T}^\text{STT}_\text{m}\\
\label{eq2}
\mathbf{\dot{n}} &=-\mathbf{m}\times \left(\gamma\mathbf{H}_\text{n}-\alpha _\text{n}\mathbf{\dot{n}}\right)-\mathbf{n}\times \left(\gamma\mathbf{H}_\text{m}-\alpha _\text{m}\mathbf{\dot{m}}\right)+\mathbf{T}^\text{STT}_\text{n}
\end{align}

where, $\gamma$ is the gyromagnetic ratio and $\lbrace\alpha_\text{m}, \alpha_\text{n}\rbrace = \lbrace \frac{1}{2}(\alpha+\alpha_c), \frac{1}{2}(\alpha-\alpha_c)\rbrace$ are the eigenvalues of the dissipation matrix $\mathcal{R}$ , which for a two sub-lattice AFM system can be written as \cite{yuan3}
\begin{equation}
\label{eq3}
\mathcal{R} = \left(
\begin{array}{cc}
 \alpha  & \alpha_c \\
 \alpha_c & \alpha \\
\end{array}\right)
\end{equation} 
where, $\alpha$ and $\alpha_c$ satisfy the condition $\alpha > \alpha_c > 0$ are damping coefficients of the system.

The effective fields $\mathbf{H}_\text{m}$ and $\mathbf{H}_\text{n}$ appearing in Eq.(\ref{eq1}) can be obtained from $\mathcal{H}_{\text{eff}}$ of the system \cite{gomonay2}
\begin{equation}
\label{eq4}
\mathbf{H}_\text{m} = -\frac{\delta \mathcal{H}_\text{eff}}{\delta \mathbf{m}}; \hspace{0.5cm}
\mathbf{H}_\text{n} = -\frac{\delta \mathcal{H}_\text{eff}}{\delta \mathbf{n}}
\end{equation} 

where, $\mathcal{H}_{\text{eff}}$ is the effective Hamiltonian of the system defined as \cite{li2,gomonay2,acharjee91}
\begin{multline}
\label{eq5}
\mathcal{H}_\text{eff} = \frac{K_\text{exc}}{4M_0}\mathbf{m}^2 + \frac{K_\text{an}}{M_0}
(\mathbf{n})_z^2 -\mathbf{D}_{12}.\left(\mathbf{m_1} \times \mathbf{m_2}\right) + K_\text{ext} (\hat{e}_x . \mathbf{m})
\\- K_\text{R}(1 
- \mathbf{m}.\mathbf{n}) + \mathbf{B}_\text{N}.\mathbf{n} 
+ \delta (\mathbf{k}_x \times \hat{e}_y). \bf{\sigma}
\end{multline}
where, $K_\text{exc}$ incorporate the exchange coupling between the magnetic sublattices and $K_\text{an}$ is the easy axis anisotropy of the system taken along $z$ - direction. The iDMI vector $\mathbf{D}_{12}$ can be defined as  $\mathbf{D}_{12} = K_\text{D} (\hat{e}_z \times \hat{e}_{12}) = K_\text{D} \hat{e}_d$, where, $K_\text{D}$ is the strength of iDMI and $\hat{e}_{12}$ is the unit vector between spins \cite{li2}. 
\begin{figure*}[hbt]
\centerline
\centerline{
\includegraphics[scale = 0.58]{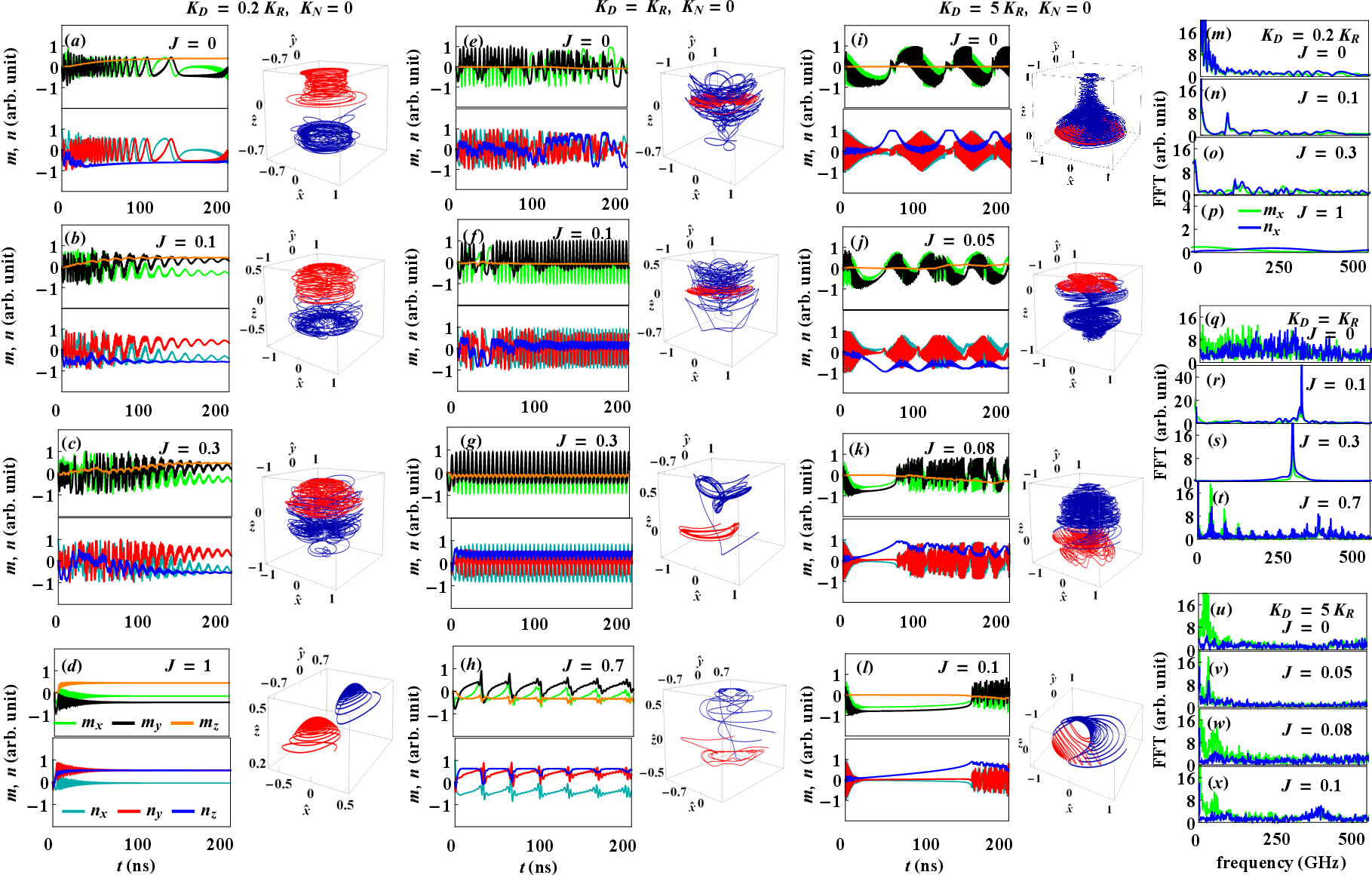}
}
\caption{Oscillation trajectories of FM ($m_x, m_y, m_z$) and AFM ($n_x, n_y, n_z$) for  
$\delta = 1 \times 10^{-10}$ eV.m, $K_{\text{ext}} = 0$ and $K_\textbf{N} = 0$ considering different choices of $J$ with (a) - (d) $K_\text{D}$ = $0.2 K_\text{R}$ (left), (e) - (h) 
$K_\text{D}$ = $K_\text{R}$ (middle) and (i) - (l) 
$K_\text{D}$ = $5 K_\text{R}$ (right). The Fourier transform of magnetization for FM and AFM are shown in plots for (m) - (p) $K_\text{D}$ = $0.2 K_\text{R}$, (q) - (t) $K_\text{D}$ = $K_\text{R}$ and (u) - (x) $K_\text{D}$ = $5 K_\text{R}$ respectively.  
}
\label{fig2}
\end{figure*}
Here, $K_\text{R}$ and $K_\text{ext}$ are the strength of RKKYI and applied external magnetic field respectively. The term $\mathbf{B}_\text{N}.\mathbf{n}$ arise due to the N\'eel interaction and defined as $\mathbf{B}_\text{N} = K_\text{N} (1, 1, 0)$. The last term of Eq. (\ref{eq5}) represent the SOC of the system with $\delta$ characterize the strength of SOC and $\mathbf{k}_x$ represents the momentum along $x$-direction.

The terms $\mathbf{T}^\text{STT}_\text{m}$ and $\mathbf{T}^\text{STT}_\text{n}$ of Eqs. (\ref{eq1}) and (\ref{eq2}) are the Spin transfer torque (STT) of the AFM and FM system defined as \cite{gomonay2}
\begin{align}
\label{eq6}
\mathbf{T}^\text{STT}_\text{m} &= \Omega_i \left[\lbrace\mathbf{m}\times(\mathbf{m}\times\mathbf{p}_\text{cur})\rbrace + 
\lbrace\mathbf{n}\times(\mathbf{n}\times\mathbf{p}_\text{cur})\rbrace
\right]\\
\label{eq7}
\mathbf{T}^\text{STT}_\text{n} &= \Omega_i \left[\lbrace\mathbf{m}\times(\mathbf{n}\times\mathbf{p}_\text{cur})\rbrace + 
\lbrace\mathbf{n}\times(\mathbf{m}\times\mathbf{p}_\text{cur})\rbrace
\right]
\end{align}
where, $\Omega_i = \frac{\gamma\hbar J}{2e\mu_0M_0t_i}$ with the index $i = 1, 2$ corresponds to the AFM and FM layers respectively and is measured in $Jm^5/A^4s$. Here, $J$ is the spin polarized current in the polarization direction $\mathbf{p}_\text{cur} = (1,1,0)$ and $t_i$ represents the thickness of the respective layers. Using Eqs. (\ref{eq4})-(\ref{eq7}) in Eq. (\ref{eq1}) and (\ref{eq2}), we obtained six non linear first order couple differential equations characterize the time evolution of the FM and the AFM order parameters viz., ($\dot{m}_x, \dot{m}_y, \dot{m}_z, \dot{n}_x, \dot{n}_y, \dot{n}_z$). An explicit form the order parameters are given in Appendix A.

To obtain dynamic equation we consider small macroscopic magnetization with $|\mathbf{m}|\ll|\mathbf{n}|$ of the AFM layer, thus can be excluded from Eq. (\ref{eq2}). Neglecting the torques and the dissipation terms we can write 
\begin{multline}
\label{eq8}  
\mathbf{\dot{n}} = -\frac{\gamma K_\text{exc}}{2M_0}(\mathbf{m} \times \mathbf{n}) -\gamma K_\text{ext}(\hat{e}_x \times \mathbf{n}) +\gamma K_\text{R}(\hat{e}_\text{r} \times \mathbf{n}) \\
+\gamma K_\text{D} \{\mathbf{n} \times  (\hat{e}_d  \times \mathbf{n})\}
\end{multline}
Performing cross product of $\mathbf{n}$ with Eq. (\ref{eq8}) and using the realtion $\mathbf{n} \times (\mathbf{m} \times \mathbf{n}) \approx 4M_0^2 \mathbf{m}$ we obtain
\begin{multline}
\label{eq9}  
\mathbf{m} = -\frac{1}{2M_0 \gamma K_\text{exc}}[(\mathbf{n} \times \mathbf{\dot{n}}) +\gamma K_\text{ext}\{\mathbf{n} \times(\hat{e}_x \times \mathbf{n})\}
\\-\gamma K_\text{R}\{\mathbf{n} \times(\hat{e}_\text{r} \times \mathbf{n})\} 
-\gamma K_\text{D} \mathbf{n} \times \{\mathbf{n} \times  (\hat{e}_d \times \mathbf{n})\}]
\end{multline}
So, the dynamic equation can be obtained by performing time differentiation Eq. (\ref{eq9}) 
\begin{multline}
\label{eq10}  
\mathbf{\dot{m}} = -\frac{1}{2M_0 \gamma K_\text{exc}}[(\mathbf{n} \times \mathbf{\ddot{n}}) 
+ \gamma K_\text{ext}\{2n\dot{n}\hat{e}_x-(\hat{e}_x.\dot{\mathbf{n}})\mathbf{n}
\\-(\hat{e}_x.\mathbf{n})\dot{\mathbf{n}}\}
-\gamma K_\text{R}\left\{2n\dot{n}\hat{e}_r - (\hat{e}_r.\dot{\mathbf{n}})\mathbf{n}-(\hat{e}_r.\mathbf{n})\dot{\mathbf{n}}\right\} \\
-\gamma K_\text{D}\left\{n^2(\dot{\mathbf{n}}\times \hat{e}_d) + 2n\dot{n}(\mathbf{n} \times \hat{e}_d)\right\}]
\end{multline}
\begin{figure*}[hbt]
\centerline
\centerline{
\includegraphics[scale = 0.58]{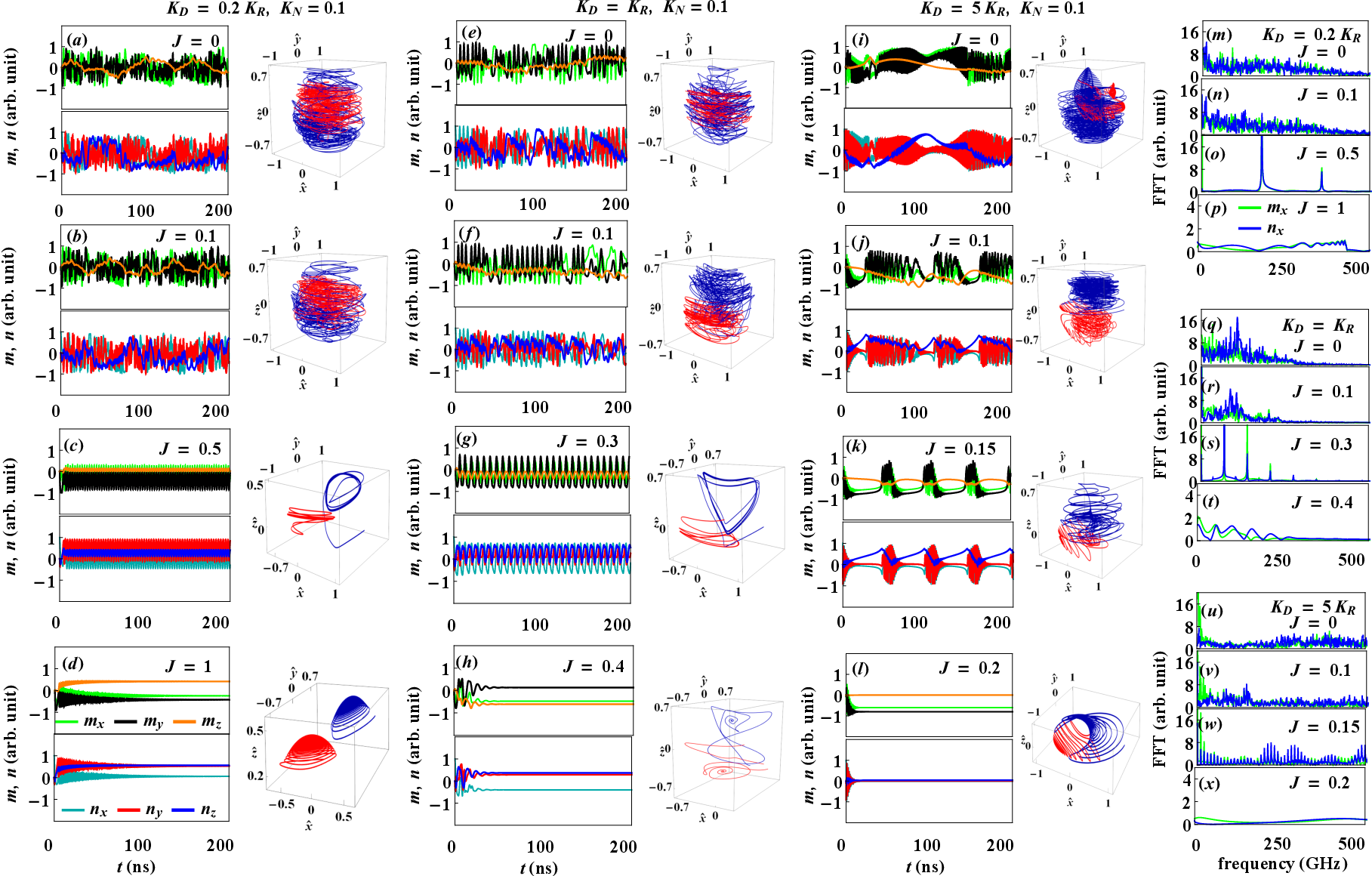}
}
\caption{Oscillation trajectories of FM ($m_x, m_y, m_z$) and AFM ($n_x, n_y, n_z$) for  
$\delta = 1 \times 10^{-10}$ eV.m, $K_{\text{ext}} = 0$ and $K_\textbf{N} = 0.1$ considering different choices of $J$ with (a) - (d) $K_\text{D}$ = $0.2 K_\text{R}$ (left), (e) - (h) 
$K_\text{D}$ = $K_\text{R}$ (middle) and (i) - (l) 
$K_\text{D}$ = $5 K_\text{R}$ (right). The Fourier transform of magnetization for FM and AFM are shown in plots for (m) - (p) $K_\text{D}$ = $0.2 K_\text{R}$, (q) - (t) $K_\text{D}$ = $K_\text{R}$ and (u) - (x) $K_\text{D}$ = $5 K_\text{R}$ respectively.  
}
\label{fig3}
\end{figure*}
\section{Results and Analysis}
\subsection{Dynamics of AFM and FM order}
We  have investigated the time evolution of the AFM ($\mathbf{n}$) and FM ($\mathbf{m}$) order by solving Eqs. (\ref{eq1}) and (\ref{eq2}) numerically. The time evolution of the AFM and FM order for different choices of $J$ in absence of N\'eel field is presented in Fig. \ref{fig2}. To further comprehend the interplay of RKKYI and iDMI on the magnetization dynamics, we consider three different scenarios of $K_\text{D}$ and $K_\text{R}$.

\subsubsection{Interplay of iDMI and RKKYI in absence of N\'eel field}
 For a system having $K_\text{D} = 0.2 K_\text{R}$, both the AFM and FM order display oscillations with different frequencies in absence of STT as illustrated in Figs. \ref{fig2}(a) and \ref{fig2}(m). The discernible decay in Fast Fourier Transform (FFT) spectra can be attributed as the interaction of the magnon with the sublattices. For $J = 0.1$ and $0.3$, both AFM and FM order undergo damped oscillations but with different frequencies as seen from Figs. \ref{fig2}(b) and \ref{fig2}(c).  The corresponding FFT in Figs. \ref{fig2}(n) and \ref{fig2}(o) further corroborate these oscillation frequencies, underlining the influence of different interactions. The damping effect arises as STT exerts an additional torque which enhances the damping of the system. With the further increase in STT ($J = 1$) the system displays rapid damped oscillations as observed from Figs. \ref{fig2}(d) and \ref{fig2}(p).

For systems with equal strength of $K_\text{D}$ and $K_\text{R}$, it is noteworthy that both AFM and FM order exhibit irregular oscillations with multiple frequencies in the absence of STT as depicted in Figs. \ref{fig2}(e) and \ref{fig2}(q). This behavior arises from the tendency of $K_\text{R}$ to facilitate a coupling between the AFM and FM order, whereas $K_\text{D}$ exerts a detrimental effect on this coupling [Ref]. The oscillations of the system tend to stabilize and display sustained oscillations with the increase in $J$ to $0.1$ and $0.3$, as evident from Figs. \ref{fig2}(f) and \ref{fig2}(g) and corresponding FFT Figs. \ref{fig2}(r) and \ref{fig2}(s) respectively. However, further escalating the value of $J$ to $0.7$, a notable transformation unfolds in both FM and AFM orders displaying harmonics, thus signaling a transition from periodic to aperiodic oscillations. 
In light of these findings, it is clear that for systems having equal strengths of $K_\text{D}$ and $K_\text{R}$, the oscillatory dynamics traverse a spectrum encompassing chaotic, highly periodic, and aperiodic motions in response to the increasing influence of the STT .
\begin{figure}[hbt]
\centerline
\centerline{
\includegraphics[scale = 0.56]{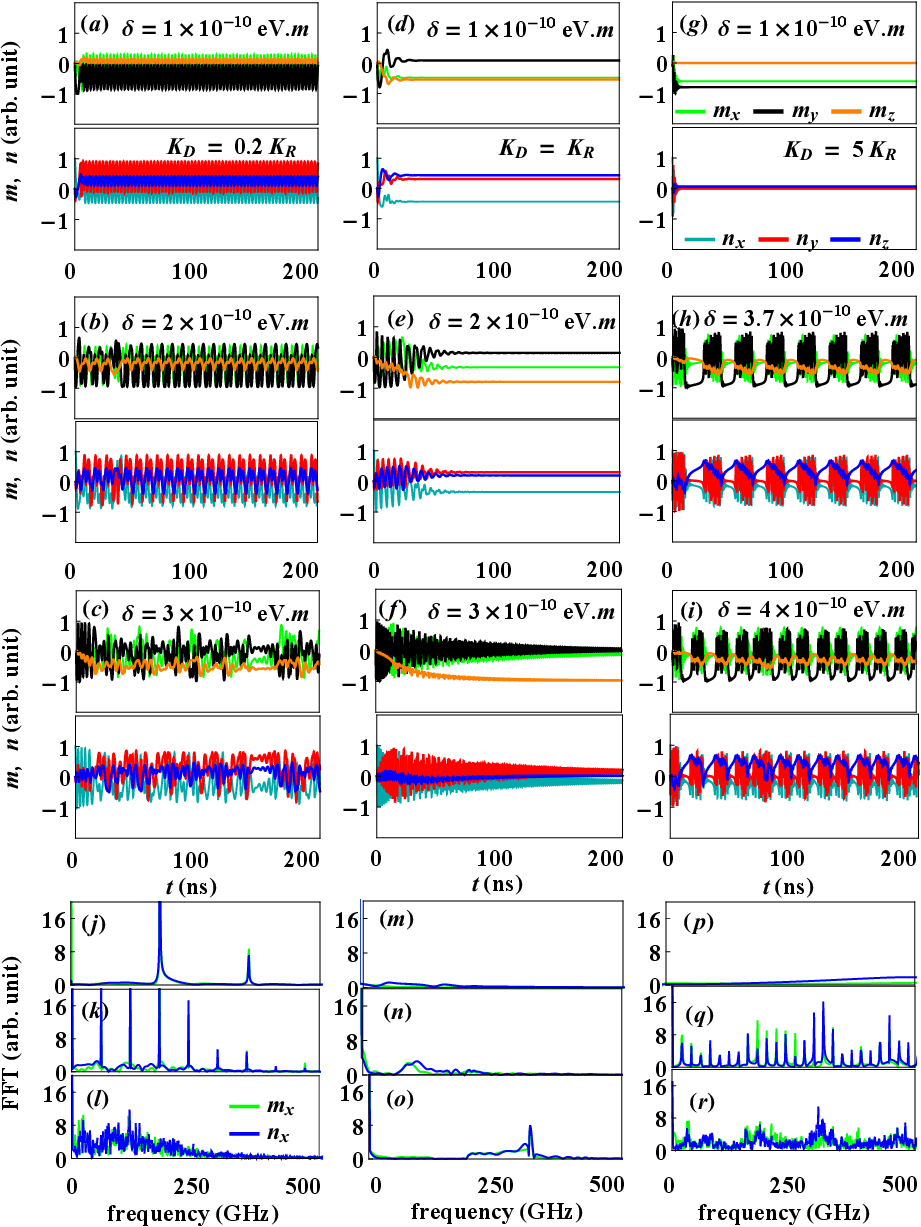}
}
\caption{Oscillation trajectories of FM ($m_x, m_y, m_z$) and AFM ($n_x, n_y, n_z$) for  
$J = 0.5$, $K_{\text{ext}} = 0$ and $K_\textbf{N} = 0.1$ considering different choices of $\delta$ with (a) - (c) $K_\text{D}$ = $0.2 K_\text{R}$ (left), (d) - (f) 
$K_\text{D}$ = $K_\text{R}$ (middle) and (g) - (i) 
$K_\text{D}$ = $5 K_\text{R}$ (right). The Fourier transform of magnetization for FM and AFM are shown in plots for (j) - (l) $K_\text{D}$ = $0.2 K_\text{R}$, (m) - (o) $K_\text{D}$ = $K_\text{R}$ and (p) - (r) $K_\text{D}$ = $5 K_\text{R}$ respectively.}
\label{fig4}
\end{figure}
In case of a system with $K_\text{D} = 5K_\text{R}$, both AFM and FM order display self-similar intermittent oscillations, primarily due to the strong $K_\text{D}$. Nonetheless, the FFT spectra reveal multiple frequency bands, indicating an inherent aperiodicity as illustrated in Figs. \ref{fig2}(i) and \ref{fig2}(u). With the increase in STT, this self-similar characteristic significantly disappears while it remains intermittent as seen from Figs. \ref{fig2}(j) -\ref{fig2}(l). However, the FFT spectra show similar characteristics in these scenarios as seen from Figs. \ref{fig2}(v) -\ref{fig2}(x). Here the system displays self-similar intermittent characteristics which are aperiodic in nature in the absence of STT. Nevertheless, as the strength of STT is enhanced, the self-similar aspect weakens while intermittent and highly aperiodic oscillations endure. Thus, our investigation reveals that, for a system with $K_\text{D} < K_\text{R}$, the system undergoes a transition from regular oscillations to highly damped oscillations. For the system, $K_\text{D}\sim K_\text{R}$, the oscillations show transition from chaotic to highly periodic behavior followed by aperiodic motion. Conversely, when $K_\text{D}$ exceeds $K_\text{R}$, the oscillations exhibit self-similar intermittent characteristics with highly aperiodic behavior.

\begin{figure}[hbt]
\centerline
\centerline{
\includegraphics[scale = 0.56]{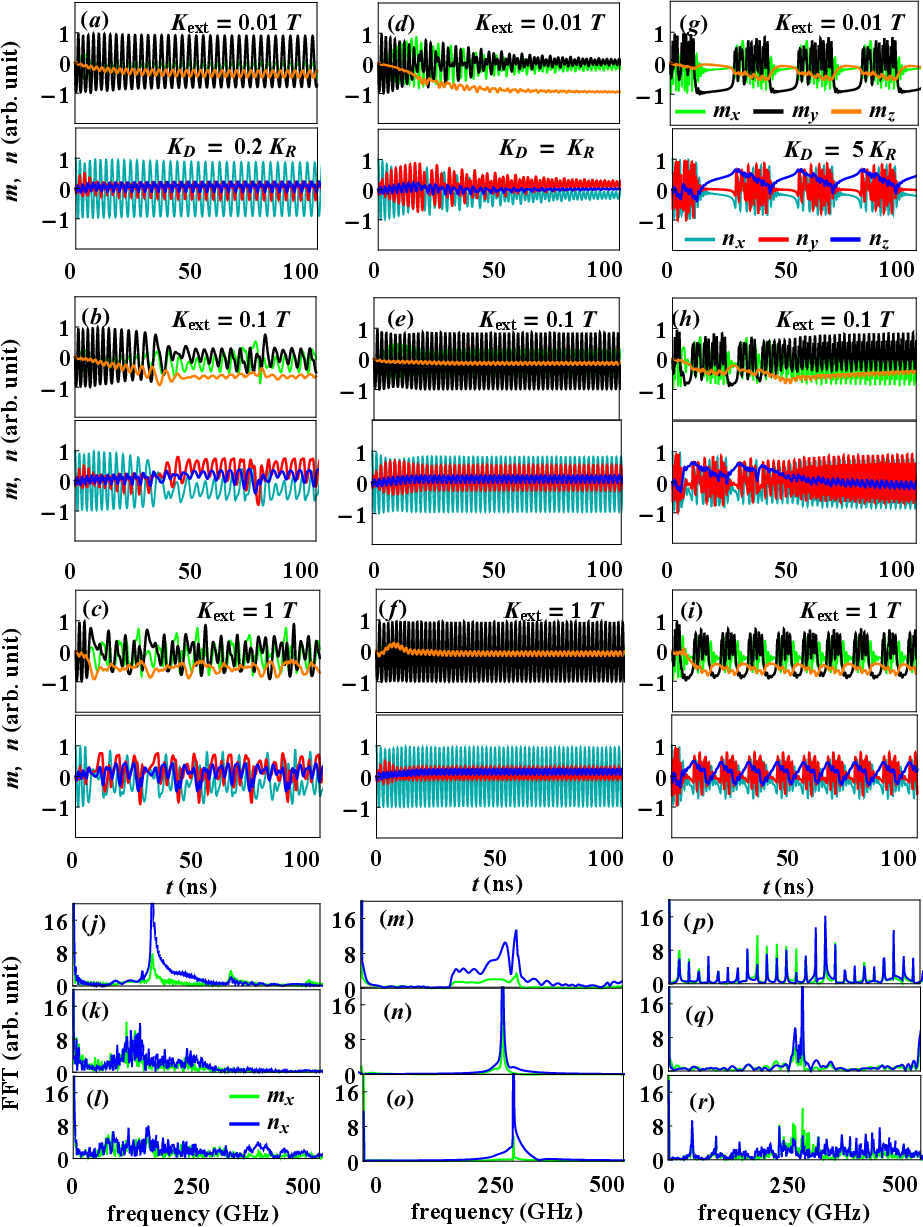}
}
\caption{Oscillation trajectories of FM ($m_x, m_y, m_z$) and AFM ($n_x, n_y, n_z$) for  
$\delta = 1 \times 10^{-10}$ eV.m, $J = 0.5$ and $K_\textbf{N} = 0.1$ considering different choices of $K_\text{ext}$ with (a) - (c) $K_\text{D}$ = $0.2 K_\text{R}$ (left), (d) - (f) 
$K_\text{D}$ = $K_\text{R}$ (middle) and (g) - (i) 
$K_\text{D}$ = $5 K_\text{R}$ (right). The Fourier transform of magnetization for FM and AFM are shown in plots for (j) - (l) $K_\text{D}$ = $0.2 K_\text{R}$, (m) - (o) $K_\text{D}$ = $K_\text{R}$ and (p) - (r) $K_\text{D}$ = $5 K_\text{R}$ respectively.
}
\label{fig5}
\end{figure}
\subsubsection{Interplay of iDMI and RKKYI in presence of N\'eel field}
To understand the role of N\'eel interaction on the AFM and FM order and its interplay with RKKYI and iDMI we consider $K_\text{N}$ = 0.1 in Fig. \ref{fig3}. For a system having $K_\text{D} < K_\text{R}$, an aperiodic oscillation with multiple frequencies are observed from Fig. \ref{fig3}(a), in the absence of STT. For a system with weak STT ($J = 0.1$), the system tends to reside in a chaotic regime. However, for a moderately strong STT ($J = 0.5$) highly periodic oscillations are observed having a major and a minor frequency as seen from Fig. \ref{fig3}(o). With further rise in STT ($J = 1$), highly damped oscillations are observed. Consequently, in the scenario where $K_\text{D} < K_\text{R}$, the system undergoes a transition from chaotic to regular behaviour as STT is increased.

A similar characteristic is observed for a system where $K_\text{D} \sim K_\text{R}$ as seen from Figs. \ref{fig3}(e) - \ref{fig3}(h). In this case, the transition occurs swiftly and can be achieved even with a low value of STT. Furthermore, multiple harmonics in the FFT spectra are observed for $J = 0.3$ as can be seen from Fig. \ref{fig3}(s). With further rise in J to $0.4$, the system exhibits rapid damping oscillation as seen from Fig. \ref{fig3}(h). This is due to the opposite but counterbalancing effect of the N\'eel field and STT.

The system having $K_\text{D} > K_\text{R}$ exhibits aperiodic oscillations in absence of STT and its corresponding FFT spectra as seen from Fig. \ref{fig3}(i) and Fig \ref{fig3}(u) respectively. The system retains its characteristics even for low values of STT as seen from Figs. \ref{fig3}(j) and \ref{fig3}(v). However, further increasing the value of $J$ to $0.15$, the system exhibits self-similar intermittent characteristics signifying regular characteristics as observed in Fig. \ref{fig3}(k) and Fig. \ref{fig3}(w). Although for slight increase in $J$ the system exhibits regular characteristics but it decays too rapidly for $J = 0.2$. In this scenario, the system is highly sensitive to STT and exhibits a transition from chaotic to regular damped behaviour. 

Our results suggest that sustained oscillations are found when iDMI is of the order of RKKYI with low values of STT and when iDMI $<$ RKKYI for moderate STT. For all other configurations, the oscillations are found to be either self-similar intermittent or damped. 
\subsubsection{Effect of Spin-Orbit Coupling}
Fig. \ref{fig4}, illustrates the impact of SOC on the time evolution of AFM and FM orders. For this analysis, we consider $J = 0.5$, $K_\text{ext} = 0$ and $K_\text{N} = 0.1$. In the case where $K_\text{D}  <  K_\text{R}$, the system exhibits oscillatory behaviour with a high degree of periodicity for $\delta = 1 \times 10^{-10}  eV.m$ as seen from Fig. \ref{fig4}(a) and its corresponding FFT in Fig. \ref{fig4}(j). However, as the strength of SOC ($\delta$) is increased to $2 \times 10^{-10} eV. m$, the oscillation of both AFM and FM orders are found to be aperiodic, manifesting multiple harmonics as evident from Figs. \ref{fig4}(b) and \ref{fig4}(k). Upon further rise in $\delta$ to $3 \times 10^{-10} eV.m$ the oscillation of both AFM and FM orders are found to be nonlinear as observed from Figs. \ref{fig4}(c) and \ref{fig4}(l). This nonlinearity arises due to the emergence of a field-like torque in presence of SOC. Consequently, in this regime, we observe a transition from linear to profoundly nonlinear behaviour with increasing SOC strength.

 The system with strength $K_\text{D} \sim K_\text{R}$ exhibit strongly damped oscillation in low SOC regime.  Moreover, the decay time significantly enhanced with the increase in SOC as observed from Figs. \ref{fig4}(d) – \ref{fig4}(f). For the region where $K_\text{D} = 5K_\text{R}$,  we observed substantially damped oscillations for weak SOC. However, self-similar quasiperiodic oscillations with multiple frequencies are emerged in Figs. \ref{fig4}(h) and \ref{fig4}(q) as $\delta \rightarrow 3.7 \times 10^{-10} eV.m$. Remarkably, for systems with $\delta = 4 \times 10^{-10} eV.m$, a highly nonlinear chaotic oscillations are noticed in Fig. \ref{fig4}(i) and in  FFT spectra Fig. \ref{fig4}(r). Thus, we observe a transition from regular to irregular behaviour with the increase in SOC. However, this characteristic is totally opposite to the system having $K_\text{D} < K_\text{R}$. This is due to the interplay of $K_\text{D}$, $K_\text{R}$ with the torque supplied by SOC. These results can be attributed as the interplay between $K_\text{D}$, $K_\text{R}$ and the torque arising due to SOC.
\subsubsection{Effect of external magnetic field}
In Fig. \ref{fig5}, we investigate the impact of external magnetic field on the dynamics of the system. We consider $J = 0.5$, $K_\text{N} = 0.1$ and $\delta = 1 \times 10^{-10} eV.m$ for this analysis. The AFM and FM orders exhibit periodic oscillations for systems having $K_\text{D} = 0.2K_\text{R}$ in presence of very low applied field as seen from Fig. \ref{fig5}(a). The existence of single major peak in the FFT spectra in Fig. \ref{fig5}(j) indicate the high periodicity of the oscillation. The periodicity of the oscillations significantly reduced with the increase in applied field to $0.1T$. This behaviour can be confirmed from the multiple bands in corresponding FFT spectra in Fig. \ref{fig5}(k).  Upon further increase in external field to $1T$, the system exhibits highly non-linear characteristics as evident from Figs. \ref{fig5}(c) and \ref{fig5}(l). This behaviour arises from the fact that the external magnetic field tends to align the magnetic moments along the x-direction.

In case of systems having $K_\text{D} \sim K_\text{R}$, we observe quasi-periodic decaying oscillations when subjected to an external field $0.01T$ as seen from Fig \ref{fig4}(f).  However, as the $K_\text{ext}$ is increased to $0.1T$, highly periodic oscillations with a single frequency are observed from Figs. \ref{fig5}(e) and \ref{fig5}(n).  This behaviour persists even with a further rise in $K_\text{ext}$ to $1T$ and it can be attributed to the minimization of damping torques in the presence of moderate and high magnetic fields.

For $K_\text{D} = 0.5 K_\text{R}$, a highly self-similar intermittent oscillations are observed when $K_\text{ext} = 0.01T$ as seen in Figs. \ref{fig5}(g) and \ref{fig5}(p). A signature of quasi-periodic oscillations are noticed in Fig. \ref{fig5}(h) and corresponding FFT spectra, Fig. \ref{fig5}(q) for $K_\text{ext} = 0.1T$. As the system is subjected to strong magnetic field $\sim 1T$, a self-similar but nonlinear oscillations are observed in Fig. \ref{fig5}(i).  In this case, both AFM and FM order exhibit multiple aperiodic frequencies as noticed from Fig. \ref{fig5}(r).  Consequently, the system undergoes a transition from aperiodic to quasiperiodic and subsequently returns to aperiodic behaviour with the increment in strength of external magnetic field.

\begin{figure*}[hbt]
\centerline
\centerline{
\includegraphics[scale = 0.63]{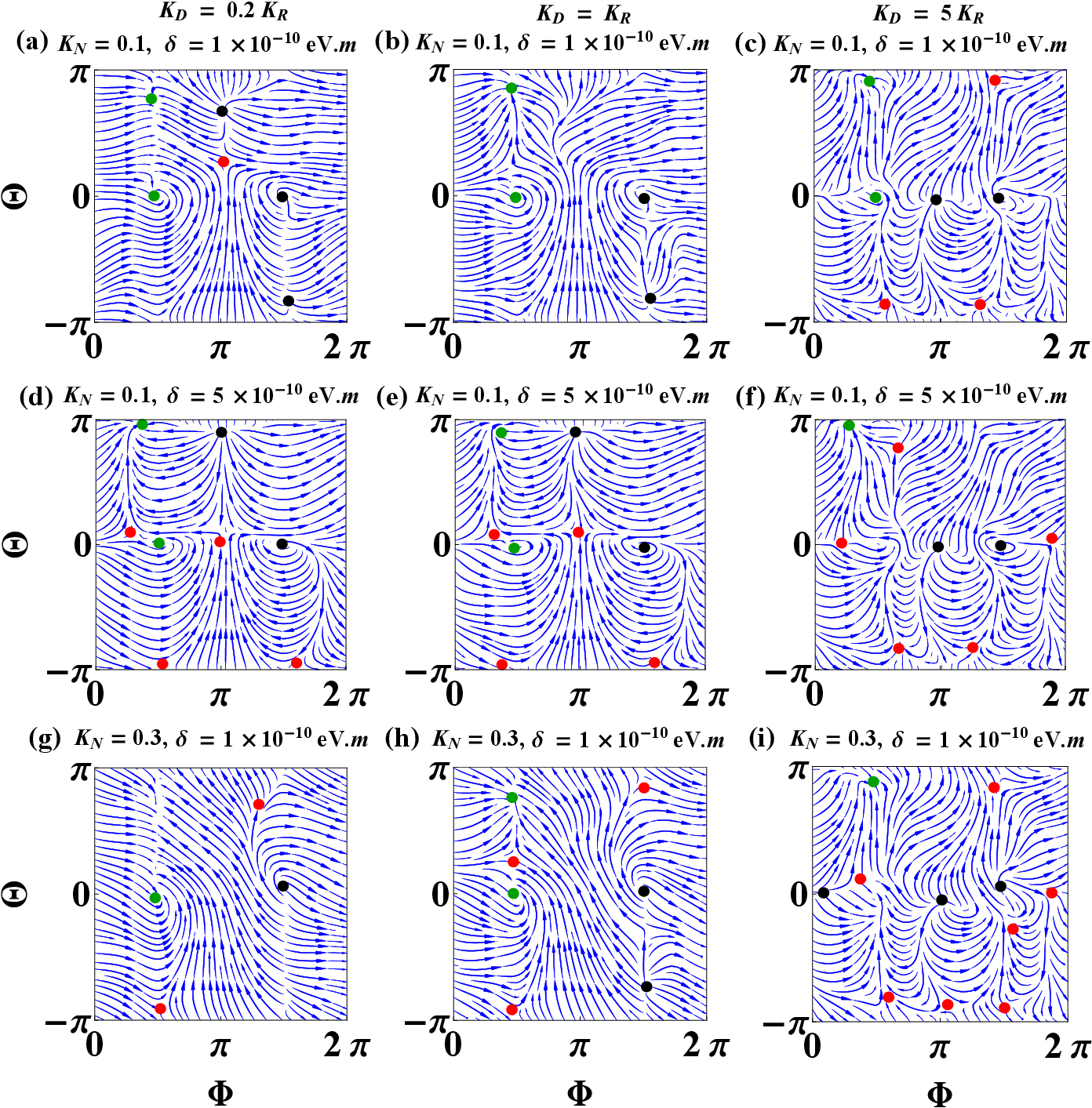}
\hspace{-0.1cm}
}
\caption{Stream plot of magnetic moments of the double barrier AFM - MTJ system. The plots in the top panel are for
$K_\text{N} = 0.1$, $\delta = 1\times 10^{-10}$ eV.m with (a) $K_\text{D} = 0.2 K_\text{R}$, (b) $K_\text{D} = K_\text{R}$ and (c) $K_\text{D} = 5 K_\text{R}$. The plots in the middle panel correspond to
$K_\text{N} = 0.1$, $\delta = 5\times 10^{-10}$ eV.m with (d) $K_\text{D} = 0.2 K_\text{R}$, (e) $K_\text{D} = K_\text{R}$ and (f) $K_\text{D} = 5 K_\text{R}$. The plots in the bottom panel are for
$K_\text{N} = 0.3$, $\delta = 1\times 10^{-10}$ eV.m with (g) $K_\text{D} = 0.2 K_\text{R}$, (h) $K_\text{D} = K_\text{R}$ and (i) $K_\text{D} = 5 K_\text{R}$ respectively. We represented stable nodes (attractor) by green dots. The black dots represent unstable nodes (repeller) while the saddle nodes are represented by red dots.}
\label{fig6}
\end{figure*}

\subsection{Equilibrium points and Stability Analysis}
Our results in Figs. \ref{fig2} - \ref{fig5}, suggest that RKKYI and iDMIs can play a significant role on magnetization dynamics of AFM system under suitable N\'eel field and SOC. The signature of transition from aperiodic to chaotic followed by regular oscillations under suitable choice of the interaction parameters, generate our interest to explore the equilibrium points of the system. Thus in Fig. \ref{fig6}, we have investigated the equilibrium points for different choices of $K_\text{D}, K_\text{R}, K_\text{N}$ and $\delta$.

Parameterizing, the AFM order parameter, $\mathbf{n} = (\sin \Theta \cos \Phi, \sin \Theta \sin \Phi, \cos \Phi)$ and using Eqs. (\ref{eq9}) and (\ref{eq10}), in Eqs. (\ref{eq1}) and (\ref{eq2}), we obtain coupled LLGS equations in polar form $(\Theta, \Phi)$, which can be expressed as
\begin{multline}
\label{eq11}  
\dot{\Theta} = \gamma ^2 \Gamma _1 K_\text{D}-\gamma ^2 \mathcal{Q}_4 \mathcal{Q}_6 K_\text{D}^2 \sin\Theta \cos\Phi
+4 \mathcal{Q}_1
\\+16\gamma  K_{\text{an}} \sin\Theta \sin 2\Phi
\end{multline}
\begin{multline}
\label{eq12}  
 \dot{\Phi} = \gamma^2  \Gamma _3 K_\text{D} \sin \Theta+8\gamma \Gamma _4-32 \gamma^2  \mathcal{Q}_2 \mathcal{Q}_5 K_\text{D}^2 \sin \Theta \
\\-2 \gamma  K_{\text{an}} \sin\Theta \cos ^2\Phi +16\mathcal{Q}_3 \cos \Phi
\end{multline}
where, $\mathcal{Q}_1, \mathcal{Q}_2, \mathcal{Q}_3, \mathcal{Q}_4, \mathcal{Q}_5, \mathcal{Q}_6$ are defined as
\begin{align}
\nonumber
\mathcal{Q}_1 &= 4 J \cos ^2\Phi -2 \sin \Theta \left(2 \gamma  \delta  \sin \Phi -J Q_5\sin \Theta\right)+\gamma  \Gamma _2\\
\nonumber
\mathcal{Q}_2 &= 4 \cos ^2\Phi-J Q_5 Q_6 \sin\Theta  \\
\nonumber
\mathcal{Q}_3 &= \sin\Theta \left(\gamma  \delta -J \sin\Theta \sin\Phi\right)+J\cos \Phi  \left(\sin ^2\Theta +1\right) \\
\nonumber
\mathcal{Q}_4 &= -4 J Q_6 \sin\Theta  \left(\sin\Phi+\cos\Phi\right)-8 \sin\Phi \\
\nonumber
\mathcal{Q}_5 &=\sin 2 \Phi -\cos 2 \Phi +1\\
\nonumber
\mathcal{Q}_6 &= \cos 2 \Theta-\cos 2 \Phi-2
\end{align}
The explicit form of the parameters $\Gamma_1, \Gamma_2, \Gamma_3$ and $\Gamma_4$ are given in Appendix B. The equilibrium points of the system are obtained by considering $(\dot{\Theta}, \dot{\Phi}) = (0, 0)$ in Eqs. (\ref{eq11}) and (\ref{eq12}). A detailed study of equilibrium points for different choices of $K_\text{D}, K_\text{R}, K_\text{N}$ and $\delta$ are shown in Table \ref{tab:my-table}.

\begin{table*}[]
\caption{\bf{Stability Analysis of AFM order}}
\label{tab:my-table}
\begin{tabular}{|c|c|c|c|c|c|}
\hline
\textbf{Point  ($\mathbf{\Theta}, \mathbf{\Phi}$)}    & $K_\text{D} : K_\text{R}$     & $K_\text{N}$       & $\mathbf{\delta} (\times 10^{-10})$ eV.m  & \textbf{Type of fixed point} & \textbf{Stability Analysis} \\ \hline
(0.5$\pi$, 0), (0.45$\pi$,0.76$\pi$)  & \multirow{3}{*}{1:5} & \multirow{3}{*}{0.1} & \multirow{3}{*}{1} & Attractor                    & Stable             \\ \cline{1-1} \cline{5-6} 
(1.5$\pi$,0), ($\pi$, 0.67$\pi$), (1.5$\pi$, -0.83$\pi$) &          &   &  & Repellor & Unstable  \\ \cline{1-1} \cline{5-6} 
($\pi$, 0.27$\pi$)                                     &                      &                    &                    & Saddle                       & Unstable           \\ \hline
(0.5$\pi$, 0), (0.45$\pi$,0.83$\pi$)                        & \multirow{2}{*}{1:1} & \multirow{2}{*}{0.1} & \multirow{2}{*}{1} & Attractor                    & Stable             \\ \cline{1-1} \cline{5-6} 
(1.5$\pi$,0), (1.5$\pi$, -0.83$\pi$)                         &                      &                    &                    & Repellor                     & Unstable           \\ \hline
(0.5$\pi$, 0), (0.45$\pi$, 0.9$\pi$)                           & \multirow{3}{*}{5:1} & \multirow{3}{*}{0.1} & \multirow{3}{*}{1} & Attractor                    & Stable             \\ \cline{1-1} \cline{5-6} 
($\pi$,0), (1.5$\pi$,0)                                 &                      &                    &                    & Repellor                     & Unstable           \\ \cline{1-1} \cline{5-6} 
(1.45$\pi$, 0.9$\pi$), (0.55$\pi$, -0.85$\pi$), (1.3$\pi$, -0.85$\pi$)                                                           &                      &                    &                    & Saddle                       & Unstable           \\ \hline
(0.5$\pi$, 0), (0.35$\pi$, 0.95$\pi$)                         & \multirow{3}{*}{1:5} & \multirow{3}{*}{0.1} & \multirow{3}{*}{5} & Attractor                    & Stable             \\ \cline{1-1} \cline{5-6} 
(1.5$\pi$, 0), ($\pi$, 0.9$\pi$)                           &                      &                    &                    & Repellor                     & Unstable           \\ \cline{1-1} \cline{5-6} 
($\pi$, 0), (0.25$\pi$, 0.1$\pi$), (0.5$\pi$, -0.95$\pi$), (1.6$\pi$, -0.95$\pi$)                                                &                      &                    &                    & Saddle                       & Unstable           \\ \hline
(0.5$\pi$, 0), (0.35$\pi$, 0.9$\pi$)                        & \multirow{3}{*}{1:1} & \multirow{3}{*}{0.1} & \multirow{3}{*}{5} & Attractor                    & Stable             \\ \cline{1-1} \cline{5-6} 
(1.5$\pi$, 0), ($\pi$, 0.9$\pi$)                             &                      &                    &                    & Repellor                     & Unstable           \\ \cline{1-1} \cline{5-6} 
($\pi$, 0.1$\pi$), (0.3$\pi$, 0.1$\pi$), (0.35$\pi$, -0.95$\pi$), (1.6$\pi$, -0.95$\pi$)                                               &                      &                    &                    & Saddle                       & Unstable           \\ \hline
(0.25$\pi$, 0.95$\pi$)                                  & \multirow{3}{*}{5:1} & \multirow{3}{*}{0.1} & \multirow{3}{*}{5} & Attractor                    & Stable             \\ \cline{1-1} \cline{5-6} 
($\pi$, 0), (1.5$\pi$, 0)                                &                      &                    &                    & Repellor                     & Unstable           \\ \cline{1-1} \cline{5-6} 
\begin{tabular}[c]{@{}c@{}}
(0.2$\pi$, 0), (1.9$\pi$, 0.05$\pi$), (0.65$\pi$, 0.75$\pi$),\\ (0.65$\pi$, -0.8$\pi$), (1.25$\pi$,   -0.8$\pi$)   \end{tabular}                            
&                      &                    &                    & Saddle                       & Unstable           \\ \hline
(0.45$\pi$, 0)                                            & \multirow{3}{*}{1:5} & \multirow{3}{*}{0.3} & \multirow{3}{*}{1} & Attractor                    & Stable             \\ \cline{1-1} \cline{5-6} 
(1.5$\pi$, 0.1$\pi$)                                    &                      &                    &                    & Repellor                     & Unstable           \\ \cline{1-1} \cline{5-6} 
(1.3$\pi$, 0.7$\pi$), (0.5$\pi$, -0.9$\pi$)                          &                      &                    &                    & Saddle                       & Unstable           \\ \hline
(0.5$\pi$, 0), (0.5$\pi$, 0.75$\pi$)                        & \multirow{3}{*}{1:1} & \multirow{3}{*}{0.3} & \multirow{3}{*}{1} & Attractor                    & Stable             \\ \cline{1-1} \cline{5-6} 
(1.5$\pi$, 0), (1.5$\pi$, -0.7$\pi$)                           &                      &                    &                    & Repellor                     & Unstable           \\ \cline{1-1} \cline{5-6} 
(0.5$\pi$, 0.25$\pi$), (0.5$\pi$, -0.9$\pi$), (1.5$\pi$, 0.8$\pi$)                                                           &                      &                    &                    & Saddle                       & Unstable           \\ \hline
(0.45$\pi$, 0.85$\pi$)                                   & \multirow{3}{*}{5:1} & \multirow{3}{*}{0.3} & \multirow{3}{*}{1} & Attractor                    & Stable             \\ \cline{1-1} \cline{5-6} 
(0.1$\pi$, 0), (-0.05$\pi$, $\pi$), (1.45$\pi$, 0.1$\pi$)                             &                      &                    &                    & Repellor                     & Unstable           \\ \cline{1-1} \cline{5-6} 
\begin{tabular}[c]{@{}c@{}}(1.85$\pi$, 0), (0.35$\pi$, 0.1$\pi$), (1.4$\pi$, 0.8$\pi$), (1.55$\pi$, -0.25$\pi$),     \\ (0.6$\pi$, -0.8$\pi$), (1.05$\pi$, -0.85$\pi$), (1.5$\pi$, -0.9$\pi$)\end{tabular} &                      &                    &                    & Saddle                       & Unstable           \\ \hline
\end{tabular}
\end{table*}

At first we set, $\delta = 1 \times 10^{-10} eV.m$ and $K_\text{N} = 0.1$. For $K_\text{D} = 0.2 K_\text{R}$, there exist two stable nodes at $ (0.5\pi, 0) $ and $(0.45\pi, 0.76\pi) $ while unstable nodes at $ (1.5\pi, 0) $, $(\pi, 0.67\pi) $, $(1.5\pi, -0.83\pi)$ and $(\pi, 0.27\pi) $ as seen from Fig. \ref{fig6}(a). But, the disappearance of the fixed points at $(\pi, 0.67\pi) $ and $(\pi, 0.27\pi) $ in Fig. \ref{fig6}(b) as $K_\text{D} \sim K_\text{R}$ indicate the presence of bifurcation in the system which can lead chaos in the system. However, the sudden appearance of a saddle nodes at $(1.45\pi, 0.9\pi), (0.55\pi, -0.85\pi)$  are observed from Fig. \ref{fig6}(c) for systems with $K_\text{D} = 5 K_\text{R}$, signifying the presence of saddle-node bifurcation in this regime. Thus the system is found to be highly sensitive to the ratio $K_\text{D}$ and $K_\text{R}$. Moreover, the sudden appearance and disappearance of the local unstable nodes pointed to the potentially chaotic nature of the oscillation of AFM order which is consistent with our findings presented in Fig. \ref{fig3}. 

Figs. \ref{fig6}(d) - \ref{fig6}(f) represents the stream plots of  $(\Theta, \Phi)$ for systems with $\delta = 5 \times 10^{-10} eV.m$ and $K_\text{N} = 0.1$. In this case the systems with $K_\text{D} <  K_\text{R}$ display regular characteristics as seen from Fig. \ref{fig6}(d). There exist two stable nodes at $(0.5\pi, 0), (0.35\pi, 0.95\pi)$ while unstable nodes at $(1.5\pi, 0), (\pi, 0.9\pi)$ and saddle nodes at $(\pi, 0), (0.25\pi, 0.1\pi), (0.5\pi, -0.95\pi), (1.6\pi, -0.95\pi)$ as represented in Table \ref{tab:my-table}. Moreover, the stream trajectories remain similar as $K_\text{D} \rightarrow K_\text{R}$ as indicated from Fig. \ref{fig6}(e). This is due to the reason that though the value of $K_\text{D}$ is increased, the strength of $K_\text{D}$ is not sufficiently strong to compensate for the torque induced by SOC resulting in regular characteristics of the oscillations.  However, a further rise in $K_\text{D}$ to $5K_\text{R}$ can result in a drastic change of the local fixed points. In this regime, the previously stable node at $(0.5\pi, 0)$ and the unstable node at $(\pi, 0)$ cease to exist entirely, and new saddle nodes emerge at $(1.9\pi, 0.05\pi)$ and $(0.65\pi, 0.75\pi)$. This transition indicates the bifurcating nature of the system and signifies a shift from regular to chaotic behavior, as vividly illustrated in Fig. 6(f).

To understand the impact of N\'eel field we set $K_\text{N} = 0.3$ with $\delta = 1 \times 10^{-10} eV.m$ in Figs. \ref{fig6}(g) - \ref{fig6}(i). For $K_\text{D} <  K_\text{R}$ , the system exhibits a stable node at $(0.45\pi, 0)$, an unstable node at $(1.5\pi, 0.1\pi)$ and two saddle nodes at $(1.3\pi, 0.7\pi), (0.5\pi, -0.9\pi)$ as seen from Fig. \ref{fig6}(g). The N\'eel field tend to align the spins along the direction of the field resulting in the disappearance of the equilibrium points with the rise in $K_\text{N}$ as depicted from Figs. \ref{fig6}(a) and \ref{fig6}(g). However, as the strength of $K_\text{D} \sim K_\text{R}$, sudden appearance of a stable node at $(0.5\pi, 0.75\pi)$, an unstable node at $(1.5\pi, -0.7\pi)$ and a saddle node at $(0.5\pi, 0.25\pi)$ is noticed from Fig. \ref{fig6}(h). This indicates the system undergoes a transition from regular to chaotic region. This is due to the fact that $K_\text{D}$ has a detrimental effect over $K_\text{N}$, and tends to destabilize the spins of the AFM system. With the further rise in the strength of $K_\text{D}$, the system becomes highly unstable resulting in appearance of multiple new saddle nodes in Fig. \ref{fig6}(i) indicating the presence of saddle node bifurcation in the system. It is due to the counterbalancing effect of  $K_\text{D}$ over $K_\text{R}$ and $K_\text{N}$. Thus, we observe that the system undergoes a transition from regular to chaotic motion with the increase in the strength of iDMI.

\subsection{Magnon dispersion relation and magnon lifetime}
\begin{figure}[hbt]
\centerline
\centerline{
\includegraphics[scale = 0.4]{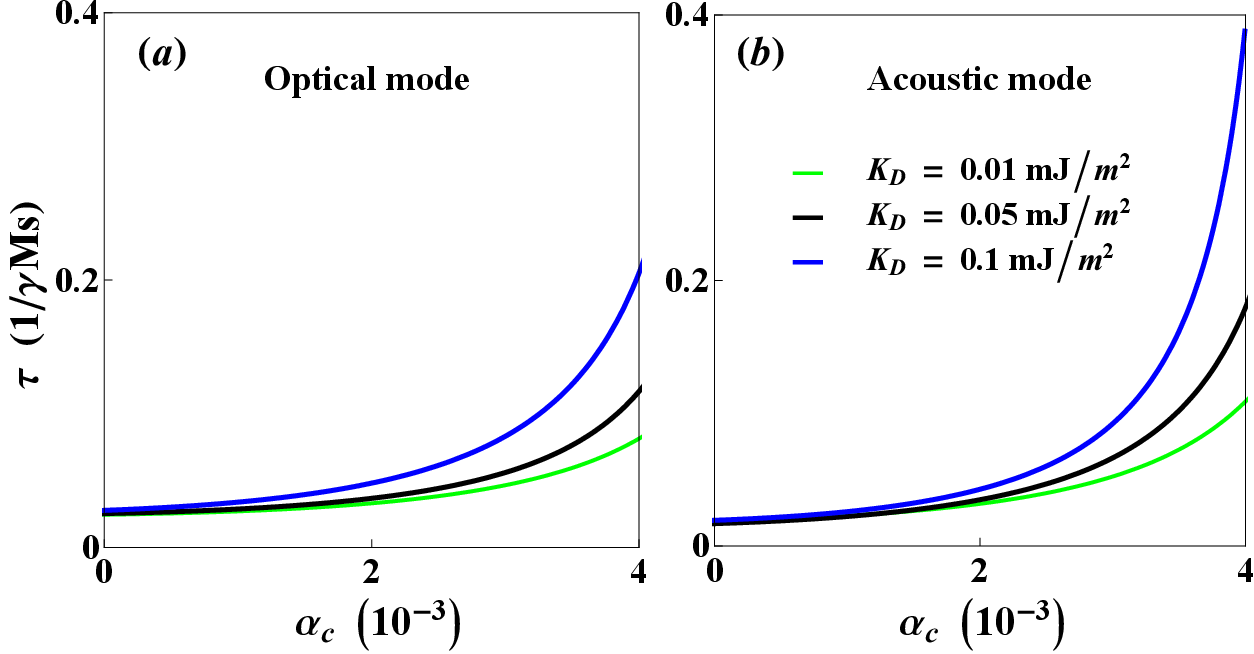}
\hspace{-0.1cm}
}
\caption{Variation of magnon lifetime with the dissipation torque $\alpha_c$ for (a) Optical mode, (b) Acoustic mode for different choices of $K_\text{D}$.}
\label{fig7}
\end{figure}

To obtain the magnon dispersion relation we consider a two sub-lattice AFM system defined by order parameters $\mathbf{m}_1$ and $\mathbf{m}_2$ respectively. In view of Eq. (\ref{eq1}), the equation of motion of two sub-lattice AFM in presence of damping and field like torques $\tau_1$ and $\tau_2$ are \cite{yuan3}
\begin{align}
\label{eq13}
\dot{\mathbf{m}}_1 &=-\gamma \left(\mathbf{m}_1 \times \mathbf{H}_1 \right) + \mathbf{m}_1 \times\left(\alpha \dot{\mathbf{m}}_1 + \alpha_c \dot{\mathbf{m}}_2\right)\\
\label{eq14}
\dot{\mathbf{m}}_2 &=-\gamma \left(\mathbf{m}_2 \times \mathbf{H}_2 \right) + \mathbf{m}_2 \times\left(\alpha \dot{\mathbf{m}}_2 + \alpha_c \dot{\mathbf{m}}_1\right)
\end{align}
 where, $\mathbf{H}_1$ and $\mathbf{H_2}$ are the effective field of sublattice 1 and 2 respectively. In order to obtain the magnon dispersion we consider a spin wave with wave vector $\mathbf{k}$ and frequency $\omega$ defined as $\mathbf{m}_j = \mathbf{m}_j^0 + \delta\mathbf{m}_j\exp\{i(\mathbf{k}.\mathbf{r} - \omega t)\}$ with, $j = 1, 2$ corresponding to the sublattice 1 and 2 respectively. Here, $\mathbf{m}_j^0$ is the ground state magnetic moment of the sublattice $j$ and $\delta\mathbf{m}_j$ is a small deviation perpendicular to $\mathbf{m}_j^0$. Following Kittel's approach, we linearized equations of Eqs. (\ref{eq13}) and (\ref{eq14}) for $\delta\mathbf{m}_1$ and $\delta\mathbf{m}_2$, the coefficient of which can be written as
\begin{equation}
\label{eq15}
\begin{matrix}
\mathcal{S} = \left(
\begin{array}{cccc}
 -i \omega  & \mathcal{R} & 0 & -i \omega  \alpha _c \\
 -\mathcal{R} & -i \omega  & i \omega  \alpha _c & 0 \\
 0 & -i \omega  \alpha _c & -i \omega  & -\mathcal{R} \\
 i \omega  \alpha _c & 0 & \mathcal{R} & -i \omega  \\
\end{array}
\right)
\end{matrix}
\end{equation}
\\ 
where, $\mathcal{R}= i \left\{\omega  \left(\alpha +\alpha _c\right)+k_z K_\text{D} \right\}+K_{\text{an}}+k_z^2 K_{\text{exc}} $
The magnon dispersion relation can be obtained by solving the secular determinant $det(\mathcal{S}) = 0$ for $\omega$.
\begin{equation}
\label{eq16}
  \left[ \omega ^2 \left\{\left(\alpha + \alpha_c\right)^2+\alpha_c^2+1\right\}-2 i \omega \eta (\alpha +\alpha_c)-\eta^2\right]^2= 0 
\end{equation}
 where, $\eta = K_{\text{an}}+i k_z K_{\text{D}} +k_z^2K_{\text{exc}}$. It is to be noted that there exist two degenerate modes in the absence of an external magnetic field. This degeneracy is due to the symmetry of the matrix $\mathcal{S}$ defined in Eq. (\ref{eq15}). The magnetization precesses circularly clockwise in one mode while counterclockwise in the other. The solutions of Eq. (\ref{eq16}) correspond to the acoustic and optical modes of magnon excitation respectively in the proposed hybrid. The magnon lifetime $\tau$ can be obtained by solving Eq. (\ref{eq16}) \cite{yuan3}
\begin{equation}
\label{eq17}
\tau  = -\frac{1}{\text{Im}(\omega)} = \frac{\alpha _c^2+\left(\alpha +\alpha _c\right){}^2+1}{ \eta\left\{\left(\alpha +\alpha _c\right)\pm\sqrt{\alpha _c^2+1}\right\}}
\end{equation}
The variation of magnon lifetime with $\alpha_c$ for both acoustic and optical modes are shown in Fig. \ref{fig7}. We consider $\alpha = 0.05$ for this analysis. The magnon lifetime for both acoustic and optical modes get enhanced with the increase in $\alpha_c$, being more prominent for low values of the conventional damping parameter $\alpha$ \cite{yuan3}. This is due to the reason that $\alpha_c$ induces a new torque acting opposite to the conventional damping torque resulting in a delay in reaching the equilibrium. Thus a dramatic increase in magnon lifetime for both optical and acoustic modes are noticed with the increase in $\alpha_c$. 

Moreover, it is of our interest to investigate the role of iDMI on the magnon lifetime. So we have plotted the lifetime for different choices of $K_{\text{D}}$ in Fig. \ref{fig7}. The lifetime for both optical and acoustic modes are found to be maximum for $K_{\text{D}} = 0.1$ $mJ/m^2$ while it is minimum for $K_{\text{D}} = 0.01$ $mJ/m^2$ with the increasing values of $\alpha_c$. This is due to the reason that the presence of iDMI induces a torque directed opposite to the conventional Gilbert damping torque resulting in further enhancement of magnon lifetime. Thus, the effect of conventional damping torque and the magnon lifetime can be tuned by controlling the iDMI of the system.  

\begin{figure*}[hbt]
\centerline
\centerline{
\includegraphics[scale = 0.5]{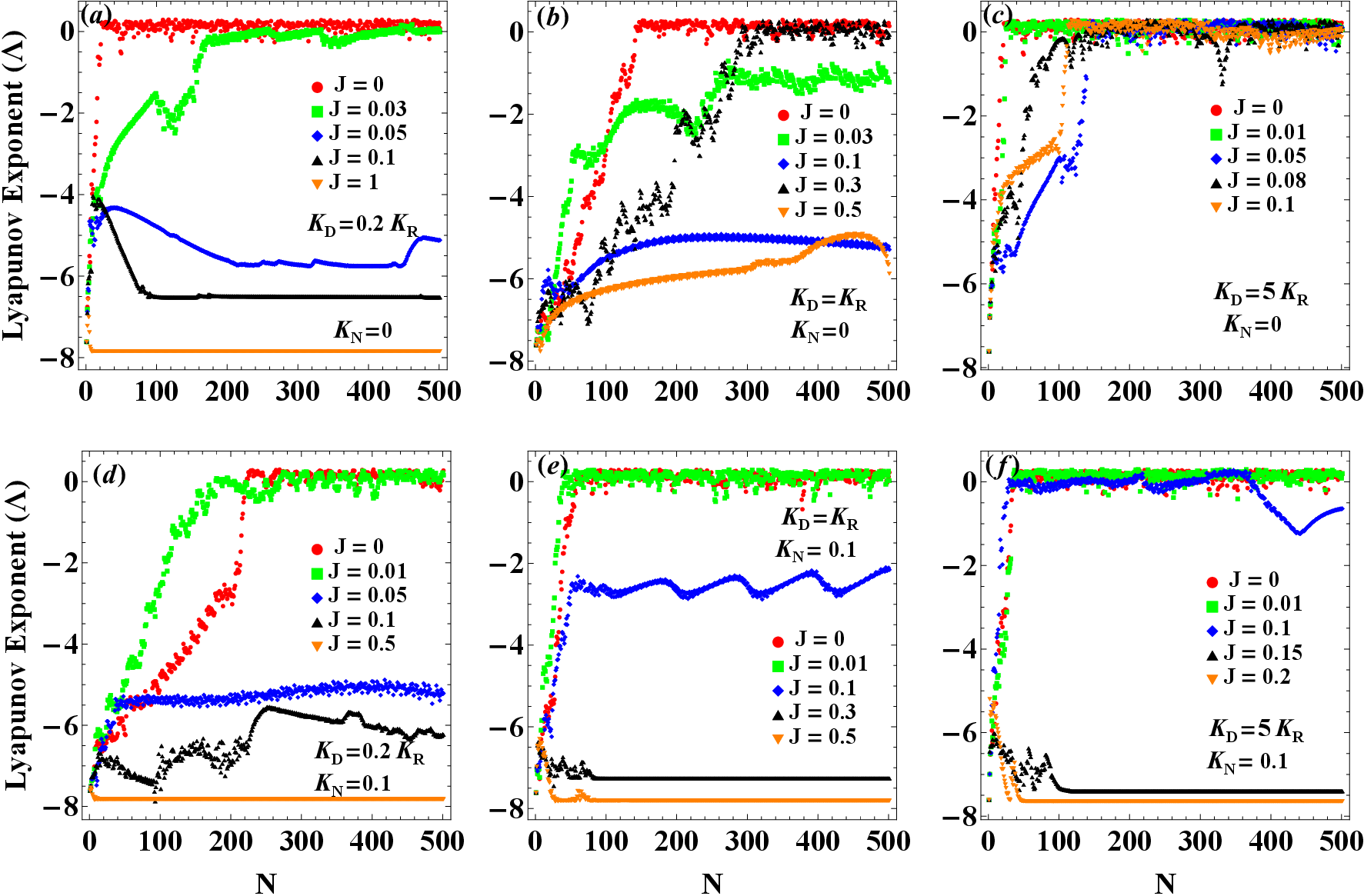}
\hspace{-0.1cm}
}
\caption{The Lyapunov exponents of the double barrier AFM - MTJ for different values of $J$. The plots in the top panel are for $K_\text{N} = 0$ with (a) $K_\text{D} = 0.2 K_\text{R}$ (left), (b) $K_\text{D} = K_\text{R}$ (middle) and (c) $K_\text{D} = 5 K_\text{R}$ (right). The plots in the bottom panel are for $K_\text{N} = 0.2$ with (e) $K_\text{D} = 0.2 K_\text{R}$ (left), (f) $K_\text{D} = K_\text{R}$ (middle) and (g) $K_\text{D} = 5 K_\text{R}$ (right).}
\label{fig8}
\end{figure*}
\subsection{Stability of the flow and Lyapunov exponents}
To comprehend the nature of magnetization dynamics we analyze the time evolution of the AFM order in Fig. \ref{fig2} - \ref{fig5}. However, to gain insight into the stability of the system we have calculated the Lyapunov exponents of the system. Lyapunov exponents are the indicators that measure the rate of convergence or divergence of the nearby trajectories. In the context of chaotic systems, a substantial increase in the rate of divergence is observed, leading to the emergence of positive Lyapunov exponents. Conversely, for regular systems, the Lyapunov exponents exhibit a negative value.

The Lyapunov exponents can be obtained by computing the average sequence of distances $\Psi_j$ (where, $j = 0,1, .... N$) between the nearest neighbor trajectories over a finite time. The underlying concept involves initiating two closely situated points with initial displacements of $ \Theta_{0,0}$ and $\Phi_{0,0}$ at time $t=0$. Thus, we can express $\Phi_{0,0} = \Theta_{0,0}+\bf{\Psi}_0$ initially at $t=0$, where $\bf{\Psi}_0$ signify the initial displacement between the two points. The points $\Theta_{0,0}$ and $\Phi_{0,0}$ evolve in accordance with the Eqs. (\ref{eq11}) and (\ref{eq12}) resulting in new positions $\Theta_{0,\mathcal{T}}$ and $\Phi_{0,\mathcal{T}}$ after a time interval $\mathcal{T}$ has transpired. Following the first time step of duration $\mathcal{T}$, we encounter the scenario where $\Phi_{0,\mathcal{T}} = \Theta_{0,\mathcal{T}} + \mathbf{\Psi}_1$, with $\mathbf{\Psi}_1 = |\mathbf{\Psi}_1|$ representing the updated distance between these two points. Treating this as the fresh reference point, we have $\Phi_{1,0} = \Theta_{1,0} + \frac{\mathbf{\Psi}_0}{\mathbf{\Psi}_1} \mathbf{\Psi}_1$, with $\Theta_{1,0} = \Theta_{0,\mathcal{T}}$. Continuing this iterative process, we generate a second set of points denoted as $\Theta_{1,\mathcal{T}}$ and $\Phi_{1,\mathcal{T}}$, where $\Phi_{1,\mathcal{T}} = \Theta_{1,\mathcal{T}} + \mathbf{\Psi}_2$. By resetting the reference points once again, we define $\Phi_{2,0} = \Theta_{2,0} + \frac{\mathbf{\Psi}_0}{\mathbf{\Psi}_2} \mathbf{\Psi}_2$, where $\Theta_{2,0} = \Theta_{1,\mathcal{T}}$, and we continue this process, evolving the new set of points for a single time step of duration $\mathcal{T}$. Repeating the steps for N times and generating a sequence of distances $\Psi_j =| \bf{\Psi}_j|$, where j = 1, 2,... N we obtain the finite time Maximum Lyapunov Exponent (MLE) which is defined as \cite{acharjee93,liu77}

\begin{equation}
\label{eq18}
\Lambda = \frac{1}{N \mathcal{T}} \sum_{j = 1}^N \ln\left(\frac{\bf{\Psi}_j}{\bf{\Psi}_0}\right)
\end{equation}
Here we consider $\mathcal{T} \ll 0$, such that $\Lambda$ is independent of $\mathcal{T}$. It is to be noted that $\lim_{N\rightarrow\infty} \Lambda \leq 0$ for regular trajectories, while $\lim_{N\rightarrow\infty} \Lambda > 0$ signify chaotic trajectories \cite{liu77}. 

From Fig. \ref{fig3}, we witness the possibility of chaotic oscillation of the AFM order under suitable choices of $K_\text{D}$ and $K_\text{R}$. Fig. \ref{fig8}, displays the MLE for the AFM system for different choices of $K_\text{N}$ and $J$. For $K_\text{D} = 0.2 K_\text{R}$ with $K_\text{N} = 0$, the MLE are found to be negative for $J \ge 0.05$ while it is found to be positive for $J < 0.05$ signifying the chaotic oscillations as seen from Fig. \ref{fig8}(a). For a system with $K_\text{D} \sim K_\text{R}$,  the MLE is found to be negative for the range $N < 180$ for $J = 0$. The MLE monotonically increases and becomes positive for $N \sim 180$ in absence of STT. For a system with $J < 0.3$, the MLE is found to be negative indicating the regular oscillation of the system. However, a totally opposite characteristics with $\Lambda > 0$ is found as $J \sim 0.3$. Moreover for higher values of $J$, a highly stable characteristic is noticed in  absence of N\'eel field. The MLEs are found to be monotonically increasing and become positive for a system with $K_\text{D} = 5 K_\text{R}$ as seen from \ref{fig8}(c). Thus we found that in the absence of N\'eel field, the system makes a transition from regular to chaotic region with the decrement in STT as the strength of iDMI approaches to RKKYI. However for the systems in which the strength of iDMI is more than the RKKYI can exhibit chaotic oscillations for all choices of STT.

Similar characteristics in MLE are also found for systems with N\'eel field strength $K_\text{N} = 0.1$ and $K_\text{D} = 0.2 K_\text{R}$. In this case, the MLE shows a linear rise for iterations $N < 200$ and becomes positive for $J \le 0.01$ while $\Lambda$ remains positive for all choices of $J > 0.01$ indicating the regular nature of the oscillations as seen from Fig. \ref{fig8}(e). A similar characteristic is also found for systems with $K_\text{D} \sim K_\text{R}$. However, in this case, the MLE becomes positive for $J \le 0.01$ under the iteration $N > 70$. A drastically different characteristic is seen in the system with $K_\text{D} = 5 K_\text{R}$ for $J = 0.15$. In this scenario, the system exhibits a transition from regular to chaotic followed by periodic oscillations as depicted in Fig. \ref{fig8}(g). Thus it can be concluded that in the presence of N\'eel field, the system exhibits chaotic oscillation under a low STT regime whilst it displays periodic oscillations in presence of high STT.

\section{Conclusions}
In summary, in this work, we have studied the spin dynamics of an SAFM$|$HM$|$ FM-based double barrier Magnetic Tunnel Junction (MTJ) in presence of Ruderman - Kittel - Kasuya - Yoside interaction (RKKYI), interfacial Dzyaloshinskii - Moriya interaction (iDMI), N\'eel field and Spin-Orbit Coupling (SOC). Moreover, we also have considered the impact of Spin Transfer Torque (STT) on the spin dynamics of the system. The oscillations of both AFM and FM orders in the proposed system are found to be strongly dependent on the strength of RKKYI and iDMI, and it exhibits a transition from regular to damped oscillations with an increase in the strength of STT for systems where iDMI is weaker than the RKKYI in absence of N\'eel field. However, the systems with equal strength of RKKY and iDMI display a transition from chaotic to periodic behaviour followed by aperiodic motion with increasing strength of STT. In contrast as the strength of iDMI exceeds the RKKYI, the system exhibits self-similar patterns indicating aperiodicity of the oscillations in the absence of the N\'eel field.
The nature of the oscillations is found to be dependent on the N\'eel field. In the presence of N\'eel field the systems with stronger RKKYI than iDMI, both AFM and FM orders exhibit chaotic oscillations for low STT but tend to display sustained oscillation for moderate strength of STT. This makes our system suitable for use in Spin Torque Oscillators. We have found similar characteristics for systems having the same strength as RKKYI and iDMI. However, the systems with iDMI $>$ RKKYI, display very sensitive dependence on the strength of STT. In this scenario, the oscillations are found to be either self-similar intermittent or damped in the presence of the N\'eel field. We found periodic sustained oscillations for low SOC but exhibited non-linearity with the increase in SOC for systems with weaker iDMI than RKKYI. The decay time was significantly enhanced with increasing SOC for systems having the same strength of iDMI and RKKYI. However, an opposite characteristic is found for iDMI $>$ RKKYI. In this case, the system undergoes regular to chaotic transition with an increase in the strength of SOC. Moreover, we found that the oscillations are tunable via a suitable external magnetic field. The oscillations are found to be periodic for low magnetic field for iDMI $<$ RKKYI, while moderate fields are favourable for sustained oscillations in systems with iDMI $\ge$ RKKYI.
 It is to be noted that the proposed system is found to exhibit saddle-node bifurcation and chaos under moderate N\'eel field and SOC with suitable iDMI and RKKYI.
Our study also reveals the impact of iDMI on the magnon lifetime. We found that the magnon lifetime for both optical and acoustic modes gets enhanced with the increase in iDMI and off-diagonal components of the damping matrix. Since iDMI and damping torques are adjustable from external sources, we can tune the lifetimes of magnons in our system by controlling these parameters. As a concluding remark, we found the oscillations are tunable via suitable choices of STT, SOC, N\'eel field and external magnetic field.  Thus, it is important to consider the impact of all the interacting variables while fabricating a magnetic tunnel junction.

\appendix
\section{Matrix representing the time derivative of FM and AFM order parameters}
An explicit form of the time derivative of the FM and AFM order parameters can be written as
\begin{widetext}
\begin{equation}
\left(
\begin{array}{c}
 \dot{m}_x \\
 \dot{m}_y \\
 \dot{m}_z \\
 \dot{n}_x \\
 \dot{n}_y \\
 \dot{n}_z \\
\end{array}
\right)= \mathcal{D}^{-1}\left(
\begin{array}{c}
 \mathcal{N}_{11} \mathcal{P}_1+\mathcal{N}_{12} \mathcal{P}_2+\mathcal{N}_{13} \mathcal{P}_3+\mathcal{N}_{14} \mathcal{P}_4+\mathcal{N}_{15} \mathcal{P}_5+\mathcal{N}_{16} \mathcal{P}_6 \\
 \mathcal{N}_{21} \mathcal{P}_1+\mathcal{N}_{22} \mathcal{P}_2+\mathcal{N}_{23} \mathcal{P}_3+\mathcal{N}_{24} \mathcal{P}_4+\mathcal{N}_{25} \mathcal{P}_5+\mathcal{N}_{26} \mathcal{P}_6 \\
 \mathcal{N}_{31} \mathcal{P}_1+\mathcal{N}_{32} \mathcal{P}_2+\mathcal{N}_{33} \mathcal{P}_3+\mathcal{N}_{34} \mathcal{P}_4+\mathcal{N}_{35} \mathcal{P}_5+\mathcal{N}_{36} \mathcal{P}_6 \\
 \mathcal{N}_{41} \mathcal{P}_1+\mathcal{N}_{42} \mathcal{P}_2+\mathcal{N}_{43} \mathcal{P}_3+\mathcal{N}_{44} \mathcal{P}_4+\mathcal{N}_{45} \mathcal{P}_5+\mathcal{N}_{46} \mathcal{P}_6 \\
 \mathcal{N}_{51} \mathcal{P}_1+\mathcal{N}_{52} \mathcal{P}_2+\mathcal{N}_{53} \mathcal{P}_3+\mathcal{N}_{54} \mathcal{P}_4+\mathcal{N}_{55} \mathcal{P}_5+\mathcal{N}_{56} \mathcal{P}_6 \\
 \mathcal{N}_{61} \mathcal{P}_1+\mathcal{N}_{62} \mathcal{P}_2+\mathcal{N}_{63} \mathcal{P}_3+\mathcal{N}_{64} \mathcal{P}_4+\mathcal{N}_{65} \mathcal{P}_5+\mathcal{N}_{66} \mathcal{P}_6 \\
\end{array}
\right)
\end{equation}
where,  $\mathcal{P}$'s and $\mathcal{N}$'s are defined as
\begin{multline}
\nonumber
\mathcal{P}_1 = -\gamma  \{n_z (K_\text{N}-K_{\text{an}} n_y)-K_\text{D} (m_z n_x-m_x n_z)-\delta n_y\}
+J (m_x m_y-m_y^2-m_z^2+n_x n_y-n_y^2-n_z^2)
\end{multline}
\begin{multline}
\nonumber
\mathcal{P}_2 = -\gamma  \{n_z (K_{\text{an}} n_x+K_\text{D} m_y-K_\text{N})-m_z (K_\text{D} n_y+K_{\text{ext}} m_x)
+\delta n_x\}+J (m_x m_y-m_x^2-m_z^2+n_x n_y-n_x^2-n_z^2)
\end{multline}
\begin{multline}
\nonumber
\mathcal{P}_3 = -\gamma (K_D \{m_z (n_x+n_y)-n_z (m_x+m_y)\}+K_{\text{ext}} m_x m_y
-K_\text{N} (n_x-n_y))
+J \{m_z (m_x+m_y)+n_z (n_x+n_y)\}
\end{multline}
\begin{multline}
\nonumber
\mathcal{P}_4 = \gamma [n_z \{m_y (K_{\text{an}}-K_{\text{exc}})+K_\text{D} (n_x-n_z)\}+m_z \{K_\text{D} (m_z
-m_x)+K_{\text{exc}} n_y-K_\text{N}\}
+K_{\text{SOC}} m_y]+J \{m_x n_y
+m_y (n_x
\\-2 n_y)-2 m_z n_z\}
\end{multline}
\begin{multline}
\nonumber
\mathcal{P}_5 = \gamma [n_z \{m_x (-K_{\text{an}}+K_{\text{exc}}+K_{\text{ext}})+K_\text{D} (n_y-n_z)\}
+m_z \{-K_\text{D} (m_y-m_z)-K_{\text{exc}} n_x+K_\text{N}\}-K_{\text{SOC}} m_x]
+J \{m_y n_x\\+m_x (n_y
-2 n_x)-2 m_z n_z\}
\end{multline}
\begin{multline}
\nonumber
\mathcal{P}_6 = \gamma [K_\text{D} \{-m_z (m_x+m_y)+m_x^2+m_y^2+n_z (n_x+n_y)-n_x^2
-n_y^2\}-m_x n_y (K_{\text{exc}}+K_{\text{ext}})
+K_{\text{exc}} m_y n_x+K_\text{N} (m_x-m_y)]
\\+J \{m_z (n_x+n_y)+n_z (m_x+m_y)\}
\end{multline}
\begin{multline}
\nonumber
\mathcal{D} = \alpha _m^3 \alpha _n \{m_x^2 (n_y^2+n_z^2)+m_z^2 (n_x^2+n_y^2)-2 m_y n_y (m_x n_x
+m_z n_z)+m_y^2 (n_x^2+n_z^2)-2 m_x m_z n_x n_z\}+\alpha _m\{\alpha _n^3 \{m_x^2 (n_y^2+n_z^2)
\\+m_z^2 (n_x^2+n_y^2)-2 m_y n_y (m_x n_x+m_z n_z)+m_y^2 (n_x^2+n_z^2)
-2 m_x m_z n_x n_z\}
+2 \alpha _n (n_x^2+n_y^2+n_z^2)\}+\alpha _m^2(\alpha _n^2 \{2 m_x^2 (m_y^2\\+m_z^2
-n_x^2)
-4 m_x n_x (m_y n_y+m_z n_z)+2 n_z^2 (-m_z^2+n_x^2+n_y^2)+2 m_y^2 (m_z
-n_y) (m_z+n_y)-4 m_y m_z n_y n_z+m_x^4+m_y^4
\\+m_z^4
+(n_x^2+n_y^2){}^2
+n_z^4\}+m_x^2+m_y^2+m_z^2)+\alpha _n^2 (m_x^2+m_y^2+m_z^2)+1
\end{multline}
\begin{multline}
\nonumber
\mathcal{N}_{11} = \alpha _m^2 [\alpha _n^2 \{m_x^2 (m_y^2+m_z^2-2 n_x^2)-2 m_x n_x (m_y n_y+m_z n_z)
+m_x^4+n_x^2 (n_x^2+n_y^2+n_z^2)\}+m_x^2\}+\alpha _m \alpha _n \{\alpha _n^2 \{m_y^2 n_x^2
\\-2 m_x m_y n_x n_y+m_x^2 n_y^2+(m_z n_x-m_x n_z){}^2]+2 n_x^2+n_y^2
+n_z^2]
+\alpha _n^2 (m_x^2+m_y^2+m_z^2)+1
\end{multline}
\begin{multline}
\nonumber
\mathcal{N}_{21} = \alpha _m[\alpha _m \{\alpha _n \{\alpha _n \{n_x n_z-m_x^2 n_x n_y
+m_x^3 m_y\}+m_z (n_x^2
+n_y^2)-n_z (m_x n_x+m_y n_y)\}+m_x m_y\}
+\alpha _n \{\alpha _n \{\alpha _n (m_z n_x
\\-m_x n_z)(m_z n_y-m_y n_z)-n_z (m_x n_x
+m_y n_y)-m_z n_z^2
+m_z (m_x^2+m_y^2+m_z^2)\}+n_x n_y\}+m_z]
\end{multline}
\begin{multline}
\nonumber
\mathcal{N}_{31} = \alpha _m[\alpha _m \{\alpha _n \{\alpha _n \{n_z \{n_x (-m_z^2+n_x^2+n_y^2)-m_x m_y n_y
-m_x^2 n_x\}+m_x m_z (m_x^2+m_y^2+m_z^2-n_x^2)-m_y m_z n_x n_y-m_x m_z n_z^2
\\+n_x n_z^3\}+n_y (m_x n_x+m_z n_z)-m_y (n_x^2+n_z^2)\}+m_x m_z\}
+\alpha _n \{\alpha _n \{\alpha _n 
(m_y n_x-m_x n_y) (m_y n_z-m_z n_y)+m_x n_x n_y
\\-m_y (m_y^2
+m_z^2-n_y^2)+m_z n_y n_z-m_x^2 m_y\}+n_x n_z\}-m_y]
\end{multline}
\begin{multline}
\nonumber
\mathcal{N}_{41} = \alpha _m (\alpha _m+\alpha _n)[\alpha _m \{m_x (m_z n_y-m_y n_z)-\alpha _n \{m_x^2 (m_y n_y
+m_z n_z)-m_x n_x (m_y^2+m_z^2+n_y^2+n_z^2)+n_x^2 (m_y n_y+m_z n_z)\}\}
\\+m_x \alpha _n (m_y n_z-m_z n_y)-m_y n_y-m_z n_z]
\end{multline}
\begin{multline}
\nonumber
\mathcal{N}_{51} = \alpha _m[\alpha _m^2 \{\alpha _n (m_x m_z^2 n_y+m_y n_x n_z^2-m_z n_z (m_x m_y+n_x n_y)
+m_x^3 n_y-m_x n_x^2 n_y+m_y n_x^3-m_x^2 m_y n_x)
\\+m_x (m_x n_z-m_z n_x)\}
+\alpha _m \{\alpha _n \{\alpha _n (m_y m_z^2 n_x+m_x n_y n_z^2-m_z n_z (m_x m_y+n_x n_y)
+m_y^3 n_x-m_x m_y^2 n_y-m_y n_x n_y^2
\\+m_x n_y^3)
+n_z (-m_z^2+n_x^2+n_y^2+n_z^2)-m_x m_z n_x-m_y m_z n_y\}+m_y n_x\}+m_x \alpha _n n_y+m_y \alpha _n^2 (m_y n_z
-m_z n_y)+n_z]
\end{multline}
\begin{multline}
\nonumber
\mathcal{N}_{61} = \alpha _m[\alpha _m^2 \{\alpha _n \{n_z \{m_x (m_y-n_x) (m_y+n_x)-m_y n_x n_y+m_x^3\}-m_z (m_x m_y+ n_y-m_x^2 n_x-n_x n_y^2-n_x^3)\}+m_x (m_y n_x\\-m_x n_y)\}+\alpha _m \{\alpha _n \{\alpha _n \{m_y^2 m_z n_x-m_y n_y (m_x m_z+n_x n_z)+m_x n_z (n_y^2+n_z^2)+m_z^3 n_x-m_x m_z^2 n_z-m_z n_x n_z^2\}
\\+m_x m_y n_x+m_y m_z n_z+m_y^2 n_y-n_y (n_x^2+n_y^2+n_z^2)\}+m_z n_x\}+m_x \alpha _n n_z+m_z \alpha _n^2 (m_y n_z-m_z n_y)-n_y]
\end{multline}
\begin{multline}
\nonumber
\mathcal{N}_{12} = \alpha _m[\alpha _m \{\alpha _n \{\alpha _n \{n_x \{-m_y m_z n_z-m_y^2 n_y+n_y (n_x^2+n_y^2+n_z^2)\}+m_x m_y (m_y^2+m_z^2-n_x^2-n_y^2)-m_x m_z n_y n_z-m_x^2 n_x n_y
\\+m_x^3 m_y\}-m_z (n_x^2+n_y^2)+n_z (m_x n_x+m_y n_y)\}+m_x m_y\}+\alpha _n \{\alpha _n 
 \{\alpha _n (m_z n_x-m_x n_z) (m_z n_y-m_y n_z)
 \\+n_z (m_x n_x
 +m_y n_y)+m_z n_z^2
-m_z (m_x^2+m_y^2+m_z^2)\}+n_x n_y\}-m_z]
\end{multline}
\begin{multline}
\nonumber
\mathcal{N}_{22} = \alpha _m[\alpha _m \{\alpha _n \{\alpha _n \{n_x \{-m_y m_z n_z-m_y^2 n_y+n_y (n_x^2+n_y^2+n_z^2)\}+m_x m_y (m_y^2+m_z^2-n_x^2-n_y^2)-m_x m_z n_y n_z-m_x^2 n_x n_y
\\+m_x^3 m_y\}-m_z (n_x^2+n_y^2)+n_z (m_x n_x+m_y n_y)\}+m_x m_y\}+\alpha _n \{\alpha _n 
 \{\alpha _n (m_z n_x-m_x n_z) (m_z n_y-m_y n_z)+n_z (m_x n_x
 \\+m_y n_y)+m_z n_z^2
-m_z (m_x^2+m_y^2+m_z^2)\}+n_x n_y\}-m_z]
\end{multline}
\begin{multline}
\nonumber
\mathcal{N}_{32} =\alpha _m(\alpha _m \{\alpha _n \{\alpha _n \{n_y n_z (-m_y^2-m_z^2+n_x^2+n_y^2)-m_x n_x (m_z n_y+m_y n_z)-m_y m_z n_z^2+m_y m_z (m_y^2+m_z^2-n_y^2)\\+m_x^2 m_y m_z +n_y n_z^3\}+m_x (n_y^2+n_z^2)-m_y n_x n_y-m_z n_x n_z\}+m_y m_z\}+\alpha _n \{\alpha _n \{\alpha _n (m_y n_x-m_x n_y) (m_z n_x-m_x n_z)\\+m_x (m_y^2+m_z^2-n_x^2)-n_x (m_y n_y+m_z n_z)+m_x^3\}+n_y n_z\}+m_x)
\end{multline}

\begin{multline}
\nonumber
\mathcal{N}_{42} =\alpha _m[\alpha _m^2 \{\alpha _n \{m_y m_z^2 n_x+m_x n_y n_z^2-m_z n_z (m_x m_y+n_x n_y)+m_y^3 n_x-m_x m_y^2 n_y-m_y n_x n_y^2+m_x n_y^3\}\\+m_y (m_z n_y-m_y n_z)\}+\alpha _m\{\alpha _n \{\alpha _n \{m_x m_z^2 n_y+m_y n_x n_z^2-m_z n_z (m_x m_y+n_x n_y)+m_x^3 n_y-m_x n_x^2 n_y+m_y n_x^3
\\-m_x^2 m_y n_x\}-n_z (-m_z^2+n_x^2+n_y^2+n_z^2)+m_x m_z n_x+m_y m_z n_y\}+m_x n_y\}+m_y \alpha _n n_x-m_xn_z \alpha _n^2 (m_z n_x-m_x n_z)]
\end{multline}

\begin{multline}
\nonumber
\mathcal{N}_{52} =\alpha _m (\alpha _m+\alpha _n)\alpha _m \alpha _n \{m_y n_y (m_z^2+n_x^2)+m_x^2 m_y n_y-m_x n_x (m_y^2+n_y^2)+m_y n_y n_z^2-m_z n_z (m_y^2+n_y^2)\}-m_y m_z n_x\\+m_x m_y n_z\}+m_y \alpha _n (m_z n_x-m_x n_z)-m_x n_x-m_z n_z\}]
\end{multline}

\begin{multline}
\nonumber
\mathcal{N}_{62} =
\alpha _m[\alpha _m^2 \{\alpha _n \{-m_x m_y m_z n_x+m_z n_y (-m_y^2+n_x^2+n_y^2)+n_z (-m_x n_x n_y-m_y n_y^2+m_x^2 m_y+m_y^3)\}+m_y (m_y n_x-m_x n_y)\}\\+\alpha _m \{\alpha _n \{\alpha _n \{m_x^2 m_z n_y-m_x n_x (m_y m_z+n_y n_z)+m_y n_z (n_x^2+n_z^2)+m_z^3 n_y-m_y m_z^2 n_z-m_z n_y n_z^2\}+m_x (-(m_y n_y\\+m_z n_z))-m_x^2 n_x+n_x (n_x^2+n_y^2+n_z^2)\}+m_z n_y\}+m_z \alpha _n^2 (m_z n_x-m_x n_z)+m_y \alpha _n n_z+n_x]
\end{multline}

\begin{multline}
\nonumber
\mathcal{N}_{13} =
\alpha _m[\alpha _m \{\alpha _n \{\alpha _n \{n_z \{n_x (-m_z^2+n_x^2+n_y^2)+m_x (-m_y) n_y-m_x^2 n_x\}+m_x m_z (m_x^2+m_y^2+m_z^2-n_x^2)-m_y m_z n_x n_y\\-m_x m_z n_z^2+n_x n_z^3\}-n_y (m_x n_x+m_z n_z)+m_y (n_x^2+n_z^2)\}+m_x m_z\}+\alpha _n \{\alpha _n \{\alpha _n (m_y n_x-m_x n_y) (m_y n_z-m_z n_y)\\-m_x n_x n_y+m_y (m_z-n_y) (m_z+n_y)-m_z n_y n_z+m_x^2 m_y+m_y^3\}+n_x n_z\}+m_y]
\end{multline}

\begin{multline}
\nonumber
\mathcal{N}_{23} =
\alpha _m [\alpha _n \{\alpha _n \{\alpha _n (m_y n_x-m_x n_y) (m_z n_x-m_x n_z)-m_x (m_y^2+m_z^2-n_x^2)+n_x (m_y n_y+m_z n_z)-m_x^3\}+n_y n_z\}-m_x\}\\+\alpha _m \{\alpha _n \{\alpha _n \{n_y n_z (-m_y^2-m_z^2+n_x^2+n_y^2)-m_x n_x (m_z n_y+m_y n_z)-m_y m_z n_z^2+m_y m_z (m_y^2+m_z^2-n_y^2)\\+m_x^2 m_y m_z+n_y n_z^3\}-m_x (n_y^2+n_z^2)+m_y n_x n_y+m_z n_x n_z\}+m_y m_z\}]
\end{multline}

\begin{multline}
\nonumber
\mathcal{N}_{33} =
\alpha _m^2 \alpha _n^2 \{n_z^2 (-2 m_z^2+n_x^2+n_y^2)-2 m_z n_z (m_x n_x+m_y n_y)+m_z^2 (m_x^2+m_y^2+m_z^2)+n_z^4\}+m_z^2\}+\alpha _m \alpha _n \{\alpha _n^2 \{m_z^2 (n_x^2+n_y^2)\\-2 m_z n_z (m_x n_x+m_y n_y)+n_z^2 (m_x^2+m_y^2)\}+n_x^2+n_y^2+2 n_z^2\}+\alpha _n^2 (m_x^2+m_y^2+m_z^2)+1
\end{multline}

\begin{multline}
\nonumber
\mathcal{N}_{43} =
\alpha _m[\alpha _m^2 \{\alpha _n \{m_y^2 m_z n_x-m_y n_y (m_x m_z+n_x n_z)+m_x n_z (n_y^2+n_z^2)+m_z^3 n_x-m_x m_z^2 n_z-m_z n_x n_z^2\}+m_z (m_z n_y\\-m_y n_z)\}+\alpha _m \{\alpha _n \{\alpha _n \{n_z \{m_x (m_y-n_x) (m_y+n_x)-m_y n_x n_y+m_x^3\}+m_z (-m_x m_y n_y-m_x^2 n_x+n_x n_y^2+n_x^3)\}\\-m_x m_y n_x-m_y m_z n_z+m_y^2 (-n_y)+n_y (n_x^2+n_y^2+n_z^2)\}+m_x n_z\}+m_x \alpha _n^2 (m_x n_y-m_y n_x)+m_z \alpha _n n_x+n_y]
\end{multline}

\begin{multline}
\nonumber
\mathcal{N}_{53} =
\alpha _m[\alpha _m^2 \{\alpha _n \{m_x^2 m_z n_y-m_x n_x (m_y m_z+n_y n_z)+m_y n_z(n_x^2+n_z^2)+m_z^3 n_y-m_y m_z^2 n_z-m_z n_y n_z^2\}+m_z (m_x n_z-m_z n_x)\}\\+\alpha _m \{\alpha _n \{\alpha _n \{-m_x m_y m_z n_x+m_z n_y (-m_y^2+n_x^2+n_y^2)+n_z(-m_x n_x n_y-m_y n_y^2+m_x^2 m_y+m_y^3)\}+m_x (m_y n_y\\+m_z n_z)+m_x^2 n_x-n_x (n_x^2+n_y^2+n_z^2)\}+m_y n_z\}+m_y \alpha _n^2 (m_x n_y-m_y n_x)+m_z \alpha _n n_y-n_x)]
\end{multline}

\begin{multline}
\nonumber
\mathcal{N}_{63} =
\alpha _m (\alpha _m+\alpha _n) \{\alpha _n \{\alpha _m \{m_z n_z (m_x^2+m_y^2+n_x^2+n_y^2)-n_z^2 (m_x n_x+m_y n_y)-m_z^2 (m_x n_x+m_y n_y)\}-m_y m_z n_x+m_x m_z n_y\}\\+\alpha _m m_z (m_y n_x-m_x n_y)-m_x n_x-m_y n_y\}
\end{multline}
\begin{multline}
\nonumber
\mathcal{N}_{14} =
\alpha _n (\alpha _m+\alpha _n) \{\alpha _m \{\alpha _n \{m_x n_x (m_y^2+m_z^2+n_y^2+n_z^2)-n_x^2 (m_y n_y+m_z n_z)-m_x^2 (m_y n_y+m_z n_z)\}-m_x m_z n_y+m_x m_y n_z\}\\+m_x \alpha _n (m_z n_y-m_y n_z)-m_y n_y-m_z n_z\}
\end{multline}
\begin{multline}
\nonumber
\mathcal{N}_{24} =
\alpha _n[\alpha _m^2 \{\alpha _n \{m_y m_z^2 n_x+m_x n_y n_z^2-m_z n_z (m_x m_y+n_x n_y)+m_y^3 n_x-m_x m_y^2 n_y-m_y n_x n_y^2+m_x n_y^3\}+m_y (m_y n_z-m_z n_y)\}\\+\alpha _m \{\alpha _n \{\alpha _n \{m_x m_z^2 n_y+m_y n_x n_z^2-m_z n_z (m_x m_y+n_x n_y)+m_x^3 n_y-m_x n_x^2 n_y+m_y n_x^3-m_x^2 m_y n_x\}+n_z (-m_z^2\\+n_x^2+n_y^2+n_z^2)-m_x m_z n_x-m_y m_z n_y\}+m_x n_y\}+m_y \alpha _n n_x+m_x \alpha _n^2 (m_x n_z-m_z n_x)+n_z]
\end{multline}
\begin{multline}
\nonumber
\mathcal{N}_{34} =
\alpha _n[\alpha _m^2 \{\alpha _n \{m_y m_z^2 n_x+m_x n_y n_z^2-m_z n_z (m_x m_y+n_x n_y)+m_y^3 n_x-m_x m_y^2 n_y-m_y n_x n_y^2+m_x n_y^3\}+m_y (m_y n_z-m_z n_y)\}\\+\alpha _m \{\alpha _n \{\alpha _n \{m_x m_z^2 n_y+m_y n_x n_z^2-m_z n_z (m_x m_y+n_x n_y)+m_x^3 n_y-m_x n_x^2 n_y+m_y n_x^3-m_x^2 m_y n_x\}+n_z (-m_z^2\\+n_x^2+n_y^2+n_z^2)-m_x m_z n_x-m_y m_z n_y\}+m_x n_y\}+m_y \alpha _n n_x+m_x \alpha _n^2(m_x n_z-m_z n_x)+n_z]
\end{multline}

\begin{multline}
\nonumber
\mathcal{N}_{44} =
\alpha _m \alpha _n \{\alpha _m^2 \{m_y^2 n_x^2-2 m_x m_y n_x n_y+m_x^2 n_y^2+(m_z n_x-m_x n_z){}^2\}+2 n_x^2+n_y^2+n_z^2\}+\alpha _n^2 \{\alpha _m^2 \{m_x^2 (m_y^2+m_z^2-2 n_x^2)\\-2 m_x n_x (m_y n_y+m_z n_z)+m_x^4+n_x^2 (n_x^2+n_y^2+n_z^2)\}+m_x^2\}+\alpha _m^2 (m_x^2+m_y^2+m_z^2)+1
\end{multline}

\begin{multline}
\nonumber
\mathcal{N}_{54} =
\alpha _n(\alpha _m \{\alpha _m \{\alpha _m (m_z n_x-m_x n_z) (m_z n_y-m_y n_z)-n_z (m_x n_x+m_y n_y)-m_z n_z^2+m_z (m_x^2+m_y^2+m_z^2)\}+n_x n_y\}+\alpha _n \\\{\alpha _m \{\alpha _m \{n_x \{m_y (-m_z) n_z-m_y^2 n_y+n_y (n_x^2+n_y^2+n_z^2)\}+m_x m_y (m_y^2+m_z^2-n_x^2-n_y^2)-m_x m_z n_y n_z-m_x^2 n_x n_y\\+m_x^3 m_y\}+m_z (n_x^2+n_y^2)-n_z (m_x n_x+m_y n_y)\}+m_x m_y\}+m_z)
\end{multline}

\begin{multline}
\nonumber
\mathcal{N}_{64} =
\alpha _n(\alpha _m \{\alpha _m \{\alpha _m (m_y n_x-m_x n_y) (m_y n_z-m_z n_y)+m_x n_x n_y-m_y (m_y^2+m_z^2-n_y^2)+m_z n_y n_z-m_x^2 m_y\}+n_x n_z\}+\alpha _n \\\{\alpha _m \{\alpha _m \{n_z \{n_x (-m_z^2+n_x^2+n_y^2)+m_x (-m_y) n_y-m_x^2 n_x\}+m_x m_z (m_x^2+m_y^2+m_z^2-n_x^2)-m_y m_z n_x n_y\\-m_x m_z n_z^2+n_x n_z^3\}+n_y (m_x n_x+m_z n_z)-m_y (n_x^2+n_z^2)\}+m_x m_z\}-m_y)
\end{multline}

\begin{multline}
\nonumber
\mathcal{N}_{15} =
\alpha _n(\alpha _m^2 \{\alpha _n \{m_x m_z^2 n_y+m_y n_x n_z^2-m_z n_z (m_x m_y+n_x n_y)+m_x^3 n_y-m_x n_x^2 n_y+m_y n_x^3-m_x^2 m_y n_x\}+m_x (m_z n_x\\-m_x n_z)\}+\alpha _m \{\alpha _n \{\alpha _n \{m_y m_z^2 n_x+m_x n_y n_z^2-m_z n_z (m_x m_y+n_x n_y)+m_y^3 n_x-m_x m_y^2 n_y-m_y n_x n_y^2+m_x n_y^3\}\\-n_z (-m_z^2+n_x^2+n_y^2+n_z^2)+m_x m_z n_x+m_y m_z n_y\}+m_y n_x\}+m_x \alpha _n n_y+m_y \alpha _n^2 (m_z n_y-m_y n_z)-n_z)
\end{multline}

\begin{multline}
\nonumber
\mathcal{N}_{25} =
\alpha _n (\alpha _m+\alpha _n) \{\alpha _m \{\alpha _n \{m_y n_y (m_z^2+n_x^2)+m_x^2 m_y n_y-m_x n_x (m_y^2+n_y^2)+m_y n_y n_z^2-m_z n_z (m_y^2+n_y^2)\}+m_y (m_z n_x\\-m_x n_z)\}+m_y \alpha _n (m_x n_z-m_z n_x)-m_x n_x-m_z n_z\}
\end{multline}

\begin{multline}
\nonumber
\mathcal{N}_{35} =
\alpha _n(\alpha _m^2 \{\alpha _n \{m_x^2 m_z n_y-m_x n_x (m_y m_z+n_y n_z)+m_y n_z (n_x^2+n_z^2)+m_z^3 n_y-m_y m_z^2 n_z-m_z n_y n_z^2\}+m_z (m_z n_x-m_x n_z)\}\\+\alpha _m \{\alpha _n \{\alpha _n \{-m_x m_y m_z n_x+m_z n_y (-m_y^2+n_x^2+n_y^2)+n_z (-m_x n_x n_y-m_y n_y^2+m_x^2 m_y+m_y^3)\}+m_x (-(m_y n_y\\+m_z n_z))-m_x^2 n_x+n_x (n_x^2+n_y^2+n_z^2)\}+m_y n_z\}+m_y \alpha _n^2 (m_y n_x-m_x n_y)+m_z \alpha _n n_y+n_x)
\end{multline}

\begin{multline}
\nonumber
\mathcal{N}_{45} =
\alpha _n(\alpha _m \{\alpha _m \{\alpha _m (m_z n_x-m_x n_z) (m_z n_y-m_y n_z)+n_z (m_x n_x+m_y n_y)+m_z n_z^2-m_z (m_x^2+m_y^2+m_z^2)\}+n_x n_y\}\\+\alpha _n \{\alpha _m \{\alpha _m \{n_x \{m_y (-m_z) n_z-m_y^2 n_y+n_y (n_x^2+n_y^2+n_z^2)\}+m_x m_y (m_y^2+m_z^2-n_x^2-n_y^2)-m_x m_z n_y n_z
-m_x^2 n_x n_y\\+m_x^3 m_y\}-m_z (n_x^2+n_y^2)+n_z (m_x n_x+m_y n_y)\}+m_x m_y\}-m_z)
\end{multline}

\begin{multline}
\nonumber
\mathcal{N}_{55} =
\alpha _m \alpha _n \{\alpha _m^2 \{m_y^2 (n_x^2+n_z^2)-2 m_y n_y (m_x n_x+m_z n_z)+n_y^2 (m_x^2+m_z^2)\}+n_x^2+2 n_y^2+n_z^2\}+\alpha _n^2 \{\alpha _m^2 \{-2 m_x m_y n_x n_y\\+m_y^2 (m_z^2-2 n_y^2)-2 m_y m_z n_y n_z+m_x^2 m_y^2+m_y^4+n_y^2 (n_x^2+n_y^2+n_z^2)\}+m_y^2\}+\alpha _m^2 (m_x^2+m_y^2+m_z^2)+1
\end{multline}

\begin{multline}
\nonumber
\mathcal{N}_{65} =
\alpha _n(\alpha _m \{\alpha _m \{\alpha _m (m_y n_x-m_x n_y) (m_z n_x-m_x n_z)+m_x (m_y^2+m_z^2-n_x^2)-n_x (m_y n_y+m_z n_z)+m_x^3\}+n_y n_z\}+\alpha _n \{\alpha _m \{\alpha _m \\\{n_y n_z (-m_y^2-m_z^2+n_x^2+n_y^2)-m_x n_x (m_z n_y+m_y n_z)-m_y m_z n_z^2+m_y m_z (m_y^2+m_z^2-n_y^2)+m_x^2 m_y m_z+n_y n_z^3\}\\+m_x (n_y^2+n_z^2)-m_y n_x n_y-m_z n_x n_z\}+m_y m_z\}+m_x)
\end{multline}

\begin{multline}
\nonumber
\mathcal{N}_{16} =
\alpha _n(\alpha _m^2 \{\alpha _n \{m_z (-m_x m_y n_y-m_x^2 n_x+n_x n_y^2+n_x^3)+n_z (m_x (m_y-n_x) (m_y+n_x)-m_y n_x n_y+m_x^3)\}+m_x (m_x n_y-m_y n_x)\}\\+\alpha _m \{\alpha _n \{\alpha _n \{m_y^2 m_z n_x-m_y n_y (m_x m_z+n_x n_z)+m_x n_z (n_y^2+n_z^2)+m_z^3 n_x-m_x m_z^2 n_z-m_z n_x n_z^2\}-m_x m_y n_x\\-m_y m_z n_z+m_y^2 (-n_y)+n_y (n_x^2+n_y^2+n_z^2)\}+m_z n_x\}+m_x \alpha _n n_z+m_z \alpha _n^2 (m_z n_y-m_y n_z)+n_y)
\end{multline}

\begin{multline}
\nonumber
\mathcal{N}_{26} =
\alpha _n(\alpha _m^2 \{\alpha _n \{-m_x m_y m_z n_x+m_z n_y (-m_y^2+n_x^2+n_y^2)+n_z (-m_x n_x n_y-m_y n_y^2+m_x^2 m_y+m_y^3)\}+m_y (m_x n_y-m_y n_x)\}\\+\alpha _m \{\alpha _n \{\alpha _n \{m_x^2 m_z n_y-m_x n_x (m_y m_z+n_y n_z)+m_y n_z (n_x^2+n_z^2)+m_z^3 n_y-m_y m_z^2 n_z-m_z n_y n_z^2\}+m_x (m_y n_y\\+m_z n_z)+m_x^2 n_x-n_x (n_x^2+n_y^2+n_z^2)\}+m_z n_y\}+m_z \alpha _n^2 (m_x n_z-m_z n_x)+m_y \alpha _n n_z-n_x)
\end{multline}

\begin{multline}
\nonumber
\mathcal{N}_{36} =
\alpha _n (\alpha _m+\alpha _n)(\alpha _m \{\alpha _n \{m_z n_z (m_x^2+m_y^2+n_x^2+n_y^2)-n_z^2 (m_x n_x+m_y n_y)-m_z^2 (m_x n_x+m_y n_y)\}-m_y m_z n_x+m_x m_z n_y\}\\+m_z \alpha _n (m_y n_x-m_x n_y)-m_x n_x-m_y n_y)
\end{multline}

\begin{multline}
\nonumber
\mathcal{N}_{46} =
\alpha _n(\alpha _m \{\alpha _m \{\alpha _m (m_y n_x-m_x n_y) (m_y n_z-m_z n_y)-m_x n_x n_y+m_y (m_z-n_y) (m_z+n_y)-m_z n_y n_z+m_x^2 m_y+m_y^3\}+n_x n_z\}\\+\alpha _n \{\alpha _m \{\alpha _m \{n_z \{n_x (-m_z^2+n_x^2+n_y^2)+m_x (-m_y) n_y-m_x^2 n_x\}+m_x m_z (m_x^2+m_y^2+m_z^2-n_x^2)-m_y m_z n_x n_y\\-m_x m_z n_z^2+n_x n_z^3\}-n_y (m_x n_x+m_z n_z)+m_y (n_x^2+n_z^2)\}+m_x m_z\}+m_y)
\end{multline}

\begin{multline}
\nonumber
\mathcal{N}_{56} =
\alpha _n(\alpha _m \{\alpha _m (\alpha _m (m_y n_x-m_x n_y) (m_z n_x-m_x n_z)-m_x (m_y^2+m_z^2-n_x^2)+n_x (m_y n_y+m_z n_z)-m_x^3)+n_y n_z\}+\alpha _n \{\alpha _m \\\{\alpha _m \{n_y n_z (-m_y^2-m_z^2+n_x^2+n_y^2)-m_x n_x (m_z n_y+m_y n_z)-m_y m_z n_z^2+m_y m_z (m_y^2+m_z^2-n_y^2)+m_x^2 m_y m_z\\+n_y n_z^3\}-m_x (n_y^2+n_z^2)+m_y n_x n_y+m_z n_x n_z\}+m_y m_z\}-m_x)
\end{multline}

\begin{multline}
\nonumber
\mathcal{N}_{66} =
\alpha _m \alpha _n \{\alpha _m^2 \{m_z^2 (n_x^2+n_y^2)-2 m_z n_z (m_x n_x+m_y n_y)+n_z^2 (m_x^2+m_y^2)\}+n_x^2+n_y^2+2 n_z^2\}+\alpha _n^2 \{\alpha _m^2 \{n_z^2 (-2 m_z^2+n_x^2+n_y^2)\\-2 m_z n_z (m_x n_x+m_y n_y)+m_z^2 (m_x^2+m_y^2+m_z^2)+n_z^4\}+m_z^2\}+\alpha _m^2 (m_x^2+m_y^2+m_z^2)+1
\end{multline}

\end{widetext}

\section{Explicit form of $\Gamma_1$, $\Gamma_2$, $\Gamma_3$ and $\Gamma_4$}
The explicit form $\Gamma_1$, $\Gamma_2$, $\Gamma_3$ and $\Gamma_4$ are given below
\begin{widetext}
\begin{multline}
\nonumber
\Gamma_1 =\frac{1}{2} K_{\text{ext}} \{J \sin \Theta \{-3 \cos 3\Phi +\cos 5\Phi-2 (16 \sin 2\Phi+3 \sin 4 \Phi +27) \cos \Phi\}+8 \cos 2\Theta (\cos\Phi+\cos 3\Phi)-J \sin 5\Theta (\sin\Phi
\\+\sin 3\Phi+\cos\Phi -\cos 3\Phi)+J \sin 3\Theta (7 \sin\Phi+8 \sin 3\Phi+\sin 5\Phi+13 \cos \Phi-4 \cos 3\Phi-\cos 5\Phi)+16 \cos\Phi\}
\\-8 \sin \Theta K_\text{R} \cos ^2\Phi \{\cos 2\Theta-J \sin \Theta (2 \cos 2\Phi+5) (\sin\Phi+\cos\Phi )-J \sin 3\Theta (\sin\Phi +\cos \Phi)-2 \sin\Phi 
+\cos 2\Phi+2\}
\end{multline}
\begin{multline}
\nonumber
\Gamma _2=-4 \gamma  K_{\text{ext}} \sin \Theta K_\text{R} \cos ^2\Phi \{J \{\sin ^2\Theta (2 \sin ^2\Phi +\sin 2\Phi )+\cos ^2\Phi\}+2 \sin \Theta \sin\Phi\}+4 \cos\Phi \{\gamma  \sin\Theta K_\text{R}^2 \cos ^2\Phi \{J \sin\Theta \\
(\sin\Phi+\cos\Phi )+1\}+K_N\}+\gamma  J K_{\text{ext}}^2 \sin ^2\Theta (\sin ^2\Theta \sin ^22\Phi+4\cos\Phi \{\sin ^2\Theta \sin ^3\Phi+\cos ^3\Phi+\sin\Phi \cos^2\Phi)\}
\end{multline}
\begin{multline}
\nonumber
\Gamma _3=K_{\text{ext}} \{-32 \sin \Theta \sin\Phi \cos ^2\Phi-4 J \cos 4\Theta \sin ^2\Phi (\sin\Phi +\cos \Phi)-J \cos 2\Theta (-18 \sin\Phi-\sin 3\Phi+\sin 5\Phi-12 \cos\Phi 
\\+3 \cos 3\Phi+\cos 5\Phi)-J (19 \sin \Phi +8 \sin 3 \Phi +\sin 5\Phi +21 \cos \Phi+3 \cos 3\Phi )\}+16 K_\text{R} \cos \Phi  \{2 J \sin ^3\Theta \sin ^4\Phi 
\\+(\sin ^2\Theta
+1) \sin\Phi  \cos ^2\Phi (2 J \sin\Theta \cos\Phi+1)+2 J (\sin ^3\Theta +\sin\Theta) \sin ^2\Phi \cos ^2\Phi+\sin ^2\Theta \sin ^3\Phi (2 J \sin \Theta \cos\Phi
\\+1)
+\cos ^2\Phi\}
\end{multline}
\begin{multline}
\nonumber
\Gamma _4=\gamma  K_{\text{ext}}^2 \{2 \{-\sin \Theta  \cos ^2\Phi +J \sin ^4\Theta  \sin ^4\Phi +J (\sin ^2\Theta +1) \cos ^4\Phi +J \sin ^2\Theta  \sin \Phi  \cos ^3\Phi +J \sin ^4\Theta  \sin ^3\Phi  \cos \Phi \}
\\+J \sin ^2\Theta  \sin ^22 \Phi ]+2 \cos \Phi  \{\gamma  \sin \Theta  K_\text{R}^2 \sin \Phi  \cos \Phi  \{J \sin \Theta  (\sin \Phi +\cos \Phi )+1\}-K_\text{N}\}
-2 \gamma  K_{\text{ext}} K_\text{R} \cos \Phi  
\\ \{\cos ^2\Theta  \cos ^2\Phi 
+\sin \Theta  (2 J \sin ^2\Theta  \sin ^3\Phi 
+\sin \Theta  \sin ^2\Phi  (2 J \sin \Theta  \cos \Phi +1)+J \cos ^3\Phi +2 J \sin \Phi  \cos ^2\Phi )\}
\end{multline}
\end{widetext}


\begin{thebibliography}{99}
\bibitem{wadley}
P. Wadley, et al., Science {\bf 351}, 587–590 (2016).
\bibitem{cheng}
R. Cheng, D. Xiao, and A. Brataas, Phys. Rev. Lett. {\bf 116}, 207603 (2016).
\bibitem{kampfrath}
T. Kampfrath, et al., Nature Photon {\bf 5}, 31–34 (2011). 
\bibitem{du}
K. Du, et al. npj Quantum Mater. {\bf 8}, 17 (2023).
\bibitem{baltz}
V. Baltz, et al., Rev. Mod. Phys. {\bf 90}, 015005 (2018).
\bibitem{zelezny}
J. Zelezny, et al., Nat. Phys. {\bf 14}, 220 (2018).
\bibitem{jungwirth}
T. Jungwirth, et al. Nature Nanotech. {\bf 11}, 231–241 (2016).
\bibitem{han}
J. Han, et al.  Nat. Mater. {\bf 22}, 684–695 (2023). 
\bibitem{marti}
X. Marti, et al., Nat. Mater. {\bf 13}, 367–374 (2014).
\bibitem{park}
B. G. Park, et al.,  Nat. Mater. {\bf 10}, 347–351 (2011).
\bibitem{jungfleisch}
M. B. Jungfleisch, W. Zhang and A. Hoffmann, Phys. Lett. A {\bf 382},  865-871 (2018).
\bibitem{kim}
T. H. Kim, et al., Phys. Rev. B {\bf 104}, 054406 (2021).
\bibitem{li22}
X. Li, X. Duan, Y. G. Semenov, K. W. Kim, J. Appl. Phys. {\bf 121}, 023907 (2017).
\bibitem{dutta}
S. DuttaGupta, et al., Nat. Commun. {\bf 11}, 5715 (2020).
\bibitem{gomonay}
O. Gomonay, T. Jungwirth, and J. Sinova, Phys. Rev. Lett. {\bf 117}, 017202 (2016).
\bibitem{kosub2}
T. Kosub, et al. Nat. Commun. {\bf 8}, 13985 (2017).
\bibitem{chen}
X. Z. Chen, et al., Phys. Rev. Lett. {\bf 120}, 207204 (2018).
\bibitem{lebrun}
R. Lebrun, et. al., Nature {\bf 561}, 222 (2018).
\bibitem{olejnik}
K. Olejnik, et. al., Sci. Adv. {\bf 4}, 3566 (2018).
\bibitem{baibich}
M. N. Baibich, et. al., Phys. Rev. Lett. {\bf 61}, 2472 (1988).
\bibitem{slonczewski1}
J. C. Slonczewski, J. Magn. Magn. Mater, {\bf 159}, L1 (1996).
\bibitem{slonczewski2}
J. C. Slonczewski, Phys. Rev. B {\bf 71}, 024411 (2005).
\bibitem{berger}
L. Berger, Phys. Rev. B. {\bf 54}, 9353 (1996).
\bibitem{li2}
S. Li, et al., Nanoscale Research Letters {\bf 14}, 315 (2019).
\bibitem{acharjee91}
S. Acharjee, et al., J. Magn. Magn. Mater. {\bf 572}, 170579 (2023).
\bibitem{parkin}
S. Parkin and D. Mauri, Phys. Rev. B {\bf 44} 7131 (1991)
\bibitem{sato}
H. Sato, et al., Appl Phys Lett {\bf 101}(2), 022414 (2012).
\bibitem{choi}
J. Y. Choi, et al., Sci Rep {\bf 8}(1), 2139 (2018).
\bibitem{garzon}
E. Garz\'on, et al., Solid-State Electron. {\bf 194}, 108315 (2022).
\bibitem{lee99}
S. E. Lee, Y. Takemura and J. G. Park, Appl. Phys. Lett. {\bf 109}, 182405 (2016).
\bibitem{iwata}
J. M. Iwata-Harms, et al. Sci Rep {\bf 8}, 14409 (2018).
\bibitem{dzyaloshinsky}
I. Dzyaloshinsky, Phys Chem Solids {\bf 4}, 241 (1958).
\bibitem{moriya}
T. Moriya, Phys Rev {\bf 120}, 91 (1960).
\bibitem{pacheco}
A. F. Pacheco, et al., Nat. Mater. {\bf 18}, 679–684 (2019).
\bibitem{caretta}
L. Caretta, et al., Nat. Commun. {\bf 11}, 1090 (2020).
\bibitem{cho}
J. Cho, et al., Nat. Commun. {\bf 6}, 7635 (2015).
\bibitem{rakibul}
M. R. K. Akanda, I. J. Park, and R. K. Lake
Phys. Rev. B {\bf 102}, 224414 (2020).
\bibitem{ding}
S. Ding, et al., Phys. Rev. B {\bf 100}, 100406(R) (2019).
\bibitem{yu99}
H. Yu, J. Xiao and H. Schultheiss, Phys. Rep.
{\bf 905}, 1-59 (2021).
\bibitem{wolf}
D. Wolf, et al., Nat. Nanotechnol. {\bf 17}, 250–255 (2022).
\bibitem{park91}
T. E. Park, et al., Phys. Rev. B {\bf 103}, 104410 (2021).
\bibitem{zink}
B. R. Zink, et al., Adv. Electron.Mater.  {\bf 8}, 2200382 (2022).
\bibitem{zhao91}
W. Zhao, et al. Nanoscale Res Lett {\bf 6}, 368 (2011).
\bibitem{rozsa}
L. R\'ozsa, et al., Phys. Rev. B {\bf 100}, 064422 (2019).
\bibitem{yang91}
C. L. Yang and C. H. Lai, Sci. Rep. {\bf 11}, 15214 (2021). 
\bibitem{volvach}
I. Volvach, A.D. Kent, E.E. Fullerton, and V. Lomakin, Phys. Rev. Applied {\bf 18}, 024071 (2022).
\bibitem{zhang92}
P. Zhang, et al., Phys. Rev. Lett. {\bf 129}, 017203 (2022).
\bibitem{wu91}
H. Wu, H., et al., Nat. Commun. {\bf 13}, 1629 (2022).
\bibitem{chiang}
C. C. Chiang, et al., Phys. Rev. Lett. {\bf 123}, 227203 (2019).
\bibitem{xu51}
Z. Xu, et al., J. Appl. Phys. {\bf 133}, 153904 (2023).
\bibitem{gomonay2}
H. V. Gomonay, R. V. Kunitsyn, and V. M. Loktev
Phys. Rev. B {\bf 85}, 134446 (2012).
\bibitem{gomonay3}
O. Gomonay, et al.  Nature Phys. {\bf 14}, 213–216 (2018). 
\bibitem{yuan3}
H. Y. Yuan, et al., EPL, {\bf 126} 67006 (2019).
\bibitem{chen91}
R. Chen, et al., Nat. Commun. {\bf 12}, 3113 (2021).
\bibitem{acharjee92}
S. Acharjee and U. D. Goswami, J. Appl. Phys. {\bf 120}, 243902 (2016)
\bibitem{coelho}
A. Chavent, et al., ACS Appl. Electron. Mater. {\bf 3}, 2607-2613 (2021).
\bibitem{acharjee93}
S. Acharjee, et al., Chaos {\bf 33}, 013136 (2023).
\bibitem{liu77}
H. F. Liu, Y. Z. Yang, Z. H. Dai and Z. H. Yu, Chaos {\bf 13}, 839–844 (2003).
\end{thebibliography}
\end{document}